\documentclass[12pt]{article}
\usepackage{amssymb}
\usepackage{bbm}
\usepackage{epsfig}
\usepackage{array}
\usepackage{float}
\usepackage{dsfont}
\usepackage{amstext}
\usepackage{amsmath}
\usepackage{psfrag}
\usepackage{a4}
\usepackage{a4wide}
\usepackage{hyperref}
\usepackage{multicol} 
\usepackage{multirow}
\usepackage{subcaption}
\usepackage{ulem}
\def\be{\begin{equation}}
\def\ee{\end{equation}}
\def\bea{\begin{eqnarray}}
\def\eea{\end{eqnarray}}

\begin{document}

\begin{center}
\baselineskip 20pt 
{\Large\bf Smooth Hybrid Inflation with Low Reheat Temperature and
	Observable Gravity Waves   
	\\ \vspace{0.2cm} in $SU(5) \times U(1)_{\chi}$ Super-GUT}
\vspace{1cm}

{\large 
	Waqas Ahmed$^{a}$ \footnote{E-mail: \texttt{\href{mailto:waqasmit@nankai.edu.cn}{waqasmit@hbpu.edu.cn}}},
	Athanasios Karozas$^{b}$ \footnote{E-mail: \texttt{\href{mailto:akarozas@uoi.gr}{akarozas@uoi.gr}}},
	George K. Leontaris$^{b,c}$ \footnote{E-mail: \texttt{\href{mailto:leonta@uoi.gr}{leonta@uoi.gr}}}  
	and Umer Zubair$^{d}$ \footnote{E-mail: \texttt{\href{mailto:umer@udel.edu}{umer@udel.edu}}}
} 
\vspace{.5cm}

{\baselineskip 20pt \it
	$^{a}$ \it
	School of Mathematics and Physics, Hubei Polytechnic University, \\
	Huangshi 435003,
	China \\
	\vspace*{6pt}
	$^b$Physics Department, Theory Division, University of Ioannina,  \\
	GR-45110 Ioannina, Greece \\
		\vspace*{6pt}
	$^c$Theoretical Physics Department, CERN,\\ CH-1211 Geneva 23, Switzerland  \\

	\vspace*{6pt}
	$^d$Department of Physics and Astronomy,  \\
	University of Delaware, Newark, DE 19716, USA \\
	\vspace{2mm} }

\vspace{1cm}
\end{center}

\begin{abstract}
We realize smooth hybrid inflation in the framework of supersymmetric $SU(5) \times U(1)_{\chi}\subset SO(10)$ model which provides a natural solution to the monopole problem  appearing in the spontaneous symmetry breaking of $SU(5)$. The breaking of $U(1)_{\chi}$ symmetry leaves a residual discrete $Z_2$ symmetry, that serves as the MSSM matter parity, realizing the possibility of the lightest supersymmetric particle as a cold dark matter candidate. The $d = 5$ proton lifetime for the decay $p \rightarrow K^+ \bar{\nu}$, mediated by color-triplet Higgsinos is found to satisfy current experimental bounds if  split-high scale SUSY scenario is employed. We show that with minimal K\"ahler potential, the soft supersymmetry breaking terms play a vital r\^ole in bringing the scalar spectral index $n_s$ within the Planck's latest bounds. In a minimal K\"ahler potential setup, small values of tensor-to-scalar ratio $r \lesssim 3.5 \times 10^{-7}$ are obtained, whereas the gravitino mass turns out to be in the range that favors PeV scale SUSY but is not sufficiently high to avoid the $d = 5$ proton decay. A non-minimal K\"ahler potential including higher order corrections is required to realize successful inflation with central value of scalar spectral index $n_s = 0.9655$, large tensor modes $r \lesssim 0.056$ and a low reheat temperature $(3\times 10^{6} \lesssim T_r \lesssim 6.5 \times 10^7 )$ GeV consistent with leptogenesis and baryogenesis.
\end{abstract}

%
\section{\large{\bf Introduction}}
Among the various proposed models of cosmic inflation, supersymmetric hybrid inflation models \cite{Dvali:1994ms,Copeland:1994vg,Linde:1997sj,Pallis:2013dxa,Buchmuller:2014epa} have gained immense attention. They provide compelling framework to realize inflation within the grand unified theories (GUTs) of particle physics. Several GUTs such as $SU(5)$~\cite{Georgi:1974sy}, Flipped $SU(5)$~\cite{Barr:1981qv,Antoniadis:1987dx} and  the Pati-Salam symmetry $SU(4)_{C} \times SU(2)_L \times SU(2)_R$~\cite{Pati:1974yy,Mohapatra:1974gc} and its supersymmetric variants ~\cite{Antoniadis:1988cm} have been employed successfully to realize hybrid inflation~\cite{Jeannerot:2000sv,Pallis:2011ps}. In this work, we implement the hybrid inflation scenario in the $SU(5)$ GUT model extended by a $U(1)_{\chi}$ symmetry~\cite{apal:2019}. The  $U(1)_{\chi}$ charge assignment is such that the whole  gauge group of the model is embedded  in  $SO(10)$, i.e., $SU(5) \times U(1)_{\chi}\in SO(10)$. This model is worth exploring   due to its various attractive features  which are not present in the stardard $SU(5)$~\cite{Georgi:1974sy}. For example,  in the  $SU(5)$ model, a discrete $Z_2$ symmetry is imposed by hand to avoid rapid proton decay whereas, in $SU(5) \times U(1)_{\chi}$ this $Z_2$ symmetry  -inherited from $SO(10)$~\cite{Kibble:1982ae} -
is encoded into $U(1)_{\chi}$ and  arises naturally  
after the spontaneous breaking of this abelian factor. This $Z_2$ symmetry induces two welcome properties. First it serves as the  matter parity of the Minimal Supersymmetric Standard Model (MSSM) and secondly it ensures the existence of a stable lightest supersymmetric particle (LSP) which  can be  
 a  viable cold dark matter candidate. Furthermore, the right-handed neutrino mass in $SU(5)$ model obtained by including the singlet right-handed neutrinos is inadequate since there is no symmetry preventing the  neutrino mass to become arbitrarily large. In $SU(5) \times U(1)_{\chi}$ model however, the right-handed neutrino mass is generated by the breaking of $U(1)_{\chi}$ symmetry after one of the fields carrying $U(1)_{\chi}$ charge acquires a Vacuum Expectation Value (VEV) at an appropriately high scale.   Hence,  $SU(5) \times U(1)_{\chi}$ retains many interesting features  of the $SO(10)$ covering GUT while it relaxes some severe constraints emanating from the $SO(10)$ unification. 
 Furthermore, there are additional constraints on the Higgs spectrum  in~ $SO(10)$ inflationary scenarios~\cite{Fukuyama:2008dv,Leontaris:2016jty}
 stemming from the fact that  large Higgs representations (such as the $\underline{126}$ of $ SO(10)$ ) induce large 
 contributions to the beta-functions so that the Renormalization Group evolution  drives the gauge couplings to the non-perturbative regime~\cite{Chang:2004pb}. 
On the contrary, we will see for example that certain $SU(5)\times U(1)$ components of the $\underline{126}$ representation play an 
 	essential r\^ole, and can be incorporated in the spectrum without spoiling the perturbative unification.
 We note in passing that in string derived constructions there are alternative   mechanisms, such as fluxes and Wilson lines breaking  GUTs
such as the $SO(10)$,  and  the $SU(5) \times U(1)_{\chi}$ follows naturally as an effective field theory model~\cite{Karozas:2020zvv}.
\\

\noindent 
The breaking of $SU(5)$ gauge symmetry leads to copious production of magnetic monopoles in conflict with the cosmological observations and the breaking of $U(1)_{\chi}$ yields topologically stable cosmic strings. The cosmic strings can survive at the end of inflation depending on the breaking scale of $U(1)_{\chi}$ chosen. In order to avoid the monopole problem, we consider a  smooth variant of hybrid inflation where the gauge symmetry is broken during inflation and disastrous monopoles are inflated away. In contrast to the shifted hybrid inflation where the radiative corrections provide the necessary slope for inflation to occur, in smooth hybrid inflation this slope is already present at the classical level, and is suitable to drive inflation. The radiative corrections are therefore assumed suppressed in smooth hybrid inflation. Moreover, in this scenario inflation ends smoothly by slow roll breaking while, in the standard and its shifted variant, the end of inflation is abrupt, followed by a waterfall.

In this paper, we consider smooth hybrid inflation in the supersymmetric $SU(5) \times U(1)_{\chi}$ with $SU(5) \times U(1)_{\chi}$ symmetry broken during inflation, inflating both the monopoles and cosmic strings away. Note that in smooth hybrid inflation, the scalar spectral index $n_{\rm s}$ lies in the observed range of Planck 
	\cite{Planck:2018jri} provided the inflationary potential incorporates either the soft supersymmetry (SUSY) breaking terms, or higher-order terms in the K\"{a}hler potential \cite{bastero,Rehman:2009nq}. Without these terms, the scalar spectral index $n_{\rm s}$ lies close to 0.98 which is acceptable only if the effective number of light neutrino species are slightly greater than 3 \cite{Ade:2015lrj}.

It has been observed (see for example~\cite{Kawasaki:2003zv,Senoguz:2003zw,uzubair:2015}), that a red tilted scalar spectral index ($n_s < 1$), consistent with the Planck bounds cannot be achieved with the minimal K\"ahler potential. In the present work we show for the first time that, by taking soft SUSY contribution into account along with the supergravity (SUGRA) corrections in a minimal K\"ahler potential setup, the predictions of smooth hybrid inflation model are consistent with the Planck's latest bounds on scalar spectral index $n_s$. We also obtain a low reheat temperature $T_r \simeq 10^6$ GeV, although the tensor to scalar ratio $r$ remains extremely small. However, the value of gravitino mass $m_{3/2}$ required to avoid $d = 5$ proton decay cannot be achieved. Remarkably, by employing a non-minimal K\"ahler potential, successful inflation is realized with the central value of Planck's bound on $n_s$ and large tensor-to-scalar ratio $r$, see Ref \cite{Shafi:2010jr}. Moreover, the dangerous $d = 5$ rapid proton decay, mediated by color-triplet Higgsinos \cite{Sakai:1981pk}, is avoided for split \cite{Giudice:2004tc} and high scale SUSY \cite{Arkani-Hamed:2004ymt}. The proton lifetime for the decay $p \rightarrow K^+ \bar{\nu}$ satisfies the current Super-Kamiokande \cite{Super-Kamiokande:2016exg} experimental bounds for the range of $SU(5)$ symmetry breaking scale $M$ and gravitino mass $m_{3/2}$, obtained in this model. We have also include  the future expected sensitivity (in the absence of any proton-decay signal) of the Hyper-Kamiokande experiment~\cite{Hyper-Kamiokande:2018ofw}. The non-thermal leptogenesis yields a low reheat temperature $ (3\times 10^{6} \lesssim T_r \lesssim 7 \times 10^7 )$ GeV, avoiding the gravitino problem~\cite{Weinberg:1982zq}~\footnote{For additional  references on this issue see for example~\cite{Ellis:2020lnc}.}
 for the full range of gravitino masses. These results can be improved by using the exact calculation for $d = 5$ proton decay lifetime, which may reduce the SUSY breaking scale  ($M_{SUSY}$) avoiding the $d = 5$ proton decay in the minimal K\"ahler potential setup as well.

The future  experiments will considerably improve the  measurements of  the tensor-to-scalar ratio $r$, a canonical measure of primordial gravity waves. One of the highlights of PRISM~\cite{PRISM:2013ybg}  is to detect inflationary gravity waves with $r$ as low as $5\times10^{-4}$, and an important goal of LiteBIRD \cite{Matsumura:2013aja} is to attain a measurement of $r$ within an uncertainty of $\delta r=10^{-3}$. Future missions include PIXIE \cite{Kogut:2011xw}, which aims to measure $r < 10^{-3}$ at five standard deviations, and CORE \cite{CORE:2016ymi}, which forecasts to lower the detection limit for the tensor-to-scalar ratio down to the $10^{-3}$ level. A low reheat temperature with observable primordial gravity waves is a particular feature of the model under consideration. We obtain a tensor to scalar ratio $r$ as large as $\sim 10^{-2}$ which is measurable by current and future experiments. 

The rest of the paper is organised as follows. In section \ref{sec2} we describe the basic features of the  $SU(5) \times U(1)_{\chi}$ model including the superfields, their charge assignments, the superpotential terms constrained by a $U(1)_R$ symmetry, and proton decay constraints. In section \ref{sec3} we discuss smooth hybrid inflation while reheating and non-thermal leptogenesis discussed in section \ref{sec4}. We present the numerical results with minimal K\"ahler potential in section \ref{sec5} and non-minimal K\"ahler potential in section \ref{sec6}. Finally we summarize the results in section \ref{sum}. 

%
\section{\large{Description of the Model}} \label{sec2}%
In this section we  present the key features of the effective $SU(5) \times U(1)_{\chi}$ model  which are essential for the description of the  inflationary scenario to be implemented subsequently. We mainly 
focus on the massless spectrum  and the  properties with respect to the symmetry breaking pattern.

The MSSM matter superfields reside in the $10$, $\bar{5}$ and $1$ dimensional representations of the group $SU(5)$ and constitute the $\underline{16}$ (spinorial) representation of $SO(10)$. Their decomposition with respect to the $SU(3)\times SU(2)_L\times U(1)_Y$ gauge symmetry is 
\begin{equation}
	\label{Mspectrum} 
	\begin{split}
F_i &\equiv10_{-1}= Q(3,2)_{\frac{1}{6}}+ u^c(\overline{3},1)_{-\frac{2}{3}} + e^c (1,1)_{1}~,   \\
\bar{f}_i &\equiv \bar{5}_{+3} = d^c (\bar{3},1)_{\frac{1}{3}} + \ell(1,2)_{-\frac{1}{2}}~,    \\
\nu_{i}^{c} &\equiv 1_{-5} =  \nu^c(1,1)_{0}~,
	\end{split}
\end{equation}
where $i = 1, 2, 3$ is the generation index. The indices on the left hand side of the above decomposition  denote the $U(1)_{\chi}$ charges (in units of $2/\sqrt{10}$) of $10,\bar 5,1$ representations, whilst on the right hand side they stand for the hypercharge
assignments. Furthermore, the singlet $1_{-5}$ is identified with the right-handed neutrino superfield $\nu^c$.

\noindent 
The scalar sector of $SU(5) \times U(1)_{\chi}$ consists of the following superfields:\\ $\alpha)$ a pair of Higgs fiveplets, $h\,\equiv 5_{2}$, $\bar{h}\,\equiv \bar{5}_{-2}$, containing the electroweak Higgs doublets denoted with $h_d, h_u$ and color Higgs triplets denoted with $D_h,\bar{D}_{\bar{h}}$;\\ $\beta)$  a Higgs superfield $\Phi$  transforming according to the adjoint representation ($\Phi\, \equiv 24_{0}$) and being responsible for breaking $SU(5)$ gauge symmetry to MSSM gauge group;\\ $\gamma)$ a pair of superfields ($\chi$, $\bar{\chi}$) which trigger the breaking of $U(1)_{\chi}$ into a $Z_2$ symmetry that is precisely the MSSM matter parity;\\ $\delta)$ finally, a gauge singlet superfield $S$ is introduced whose scalar component acts as an inflaton.
\\ The decomposition of the above $SU(5)$ representations under the MSSM gauge group are
\begin{equation}
	\label{Hspectrum} 
	\begin{split}
\Phi &= \Phi_{24}(1, 1)_{0} + W_H (1, 3)_{0} + G_H (1, 8)_{0}  +  Q_H(3, 2)_{-\frac{5}{6}} + \bar{Q}_H(3, 2)_{\frac{5}{6}},  \nonumber \\ 
h &= D_h(3,1)_{-\frac{1}{3}} + h_u (1,2)_{\frac{1}{2}}~,  \nonumber \\
\bar{h} &= \bar{D}_{\bar{h}}(\bar{3},1)_{\frac{1}{3}} + h_d(1,2)_{-\frac{1}{2}},\nonumber \\
\chi &= 1_{10}, \quad \bar{\chi} = 1_{-10}, \quad S = 1_{0}~,
	\end{split}
\end{equation}
where the indices on the right hand side of the superfields $\Phi$, $h$ and $\bar{h}$ denote the hypercharge assignments, while for the superfields $\chi$, $\bar{\chi}$ and $S$, they represent the $U(1)_{\chi}$ charges.  It should be observed  that the singlets $\chi$, $\bar{\chi}$ originate from the decomposition of $126$ representation of $SO(10)$
\begin{equation}
	126 \rightarrow 1_{-10} + \overline{5}_{-2} + 10_{-6} + \overline{15}_{6} + 45_{2} + \overline{50}_{-2}.
\end{equation}
\begin{table}[!htb]
	\setlength\extrarowheight{5pt}
	\centering
	\begin{tabular}{|c||c|c|c|c|c|c|c|c|c|}
		\hline \hline
		\multirow{3}{*}{\rotatebox{90}{\textbf{Groups}}} & \multicolumn{9}{c|}{\textbf{Superfields/Representations}}                                       \\ \cline{2-10} 
		& \multicolumn{3}{c|}{\textbf{Matter sector}} & \multicolumn{6}{c|}{\textbf{Scalar sector}} \\ \cline{2-10} 
		&~ $\bar{f}_i$~          &~ $F_i$~         & ~$\nu_{i}^c$~         & ~$\Phi$~   &~ $h$~   &~ $\bar{h}$~  & ~$\chi$~  &~ $\bar{\chi}$~  &~ $S$~   \\ \hline \hline
		$SU(5)$                   &      $\bar{5}$      &     $10$      &     $1$      &  $24$   &  $5$   &  $\bar{5}$  &  $1$  &  $1$  &  $1$    \\ \hline
		$U(1)_{\chi}$                   &     $-1$       &      $3$     &     $-5$      &  $0$   &  $2$   &  $-2$  &  $10$  &  $-10$  &  $0$   \\ \hline
		$U(1)_{R}$                   &      $\frac{3}{10}$      &    $\frac{1}{10}$       &     $\frac{1}{2}$      &   $0$  &   $\frac{2}{5}$  &  $\frac{3}{5}$  &  $0$  &  $0$  &  $1$   \\ \hline
		$Z_3$                      &      $0$      &       $0$    &     $0$      &  $1$   &  $0$   &  $0$  &  $0$  &  $0$  &  $0$   \\ \hline 
		$L$                      &      $0$      &       $0$    &     $-1$      &  $0$   &  $0$   &  $0$  &  $2$  &  $-2$  &  $0$   \\ \hline
		\hline
	\end{tabular}
	\caption{The lepton number $L$, $SU(5) \times U(1)_{\chi}$, $U(1)_R$ and $Z_3$ charges of  matter and scalar superfields in smooth hybrid inflation model. The $U(1)_{\chi}$ charges are given in units of $\frac{2}{\sqrt{10}}$.}
	\label{tab:field_charges}
\end{table}
The charge assignments listed in Table \ref{tab:field_charges} may suggest that the non zero VEVs of $\chi$, $\bar{\chi}$ (carrying  $U(1)_{\chi}$ charges $10$ and $-10$ respectively) break $U(1)_{\chi}$ to $Z_{10}$. However, since the $Z_5$ subgroup of $Z_{10}$ also resides in $SU(5)$, the effective unbroken discrete symmetry is $Z_2$ \cite{apal:2019,Kibble:1982ae}. In $SO(10)$ representations, the $Z_2$ symmetry is manifested as $D$-parity, i.e., the discrete $SO(10)$ element\cite{Chang:1983fu,Kibble:1982ae} which transforms left handed fields to their right-handed conjugates. Since the matter fields contain only fields of a single chirality, there can be no well-defined notion of parity in $SO(10)$. $D$-parity  then plays a role to create the possibility for the presence of $C$ and $P$ at lower energies and acts as MSSM matter parity, see Ref \cite{Chang:1983fu, Severson:2016rqb}.  Moreover, the $Z_2$ symmetry and the particular field assignment with standard matter parity is justified by the $SO(10)$ embedding of the present model. This can be seen by noticing that the content of each fermion generation (including the right-handed neutrino)  is embedded in the fundamental 16-spinorial representation of $SO(10)$ (hence, as it was observed in~\cite{Kibble:1982ae} the actual symmetry is its simply-connected spin(10)) and the  Higgs in 10 representation of $SO(10)$.  The successive breaking with the $126_H$ (which includes the singlet $\chi$) and $10_H$ (which contains the SM Higgs), defines a $Z_2$ matter parity which remains unbroken by the 
subsequent symmetry breaking to SM group.

In the following we implement the $R$-charge assignment (as shown in Table~\ref{tab:field_charges}) of the superfields proposed in \cite{Khalil:2010cp}. In addition, imposing a $Z_{3}$ symmetry on the Higgs superfield $\Phi$ under which, $\Phi \rightarrow e^{2 \dot{\iota} \pi / 3} \Phi$ (with other fields trasforming trivially), only cubic powers of $\Phi$ are allowed in the superpotential. The complete spectrum of the model with  its various transformation properties are shown in Table~\ref{tab:field_charges}. 
The  $SU(5) \times U(1)_{\chi}$, $U(1)_R$, and $Z_{3}$ symmetric superpotential of the model with the leading-order non-renormalizable terms is given by {\footnote{For smooth SU(5) hybrid inflation, see ref.~\cite{uzubair:2015}.}}
\bea%
W &=& S\left( \mu^2 + \frac{Tr(\Phi^3)}{m_{P}}
\right) + \gamma \frac{\bar{h} \Phi^3 h}{m_{P}^{2}} + \delta \bar{h} h  + \sigma_{\chi} S \left( \chi \bar{\chi} - \mu_{\chi}^2 \right)\nonumber \\
&+& y_{ij}^{(u)}\,F_i\,F_j\,h + y_{ij}^{(d,e)}\,F_i\,\bar{f}_j\,\bar{h}
+y_{ij}^{(\nu)}\,\nu_{i}^c\,\bar{f}_j\,h + \zeta_{ij} \chi \nu_{i}^{c} \nu_{j}^{c} , \label{superpotential}
\eea %
where $\mu$ is a superheavy mass and $m_P = 2.43 \times 10^{18}$ GeV is the reduced Planck mass. 
The Yukawa couplings $y_{ij}^{(u)}$, $ y_{ij}^{(d,e)}$, $y_{ij}^{(\nu)}$ in the second line of \eqref{superpotential} generate Dirac masses for quarks and leptons after the electroweak symmetry breaking, whereas $m_{\nu_{ij}} = \zeta_{ij} \mu_{\chi}$ is the right-handed neutrino mass matrix, generated after $\chi$ acquires a VEV,  $\langle\chi\rangle = \mu_{\chi}$, breaking the $U(1)_{\chi}$ factor.  

A few comments regarding the merits of the additional symmetries are worth mentioning. By virtue of the  global $U(1)_{R}$ symmetry the superpotential $W$ exhibits a number of interesting features. First, we observe that   only linear terms in $S$ are allowed in $W$ whereas higher order ones, such as $S^2$, are prevented. This is a welcome fact since $S^2$-terms could generate an inflaton mass of Hubble size, $H \simeq \sqrt{V(x) / 3 m_P^2} $, invalidating the inflationary scenario. Furthermore, under the action of the $U(1)_{R}$ symmetry the model naturally avoids the $\eta$ problem~\cite{Linde:1997sj}, that appears when SUGRA corrections are included. Finally, due $U(1)_R$  several dangerous dimension-5 proton decay operators are suppressesed. The $Z_{3}$ symmetry plays an important r\^ole in realizing smooth hybrid inflation. At this point it should be emphasized that in the absence of this symmetry, the above superpotential \eqref{superpotential} reduces to the superpotential of shifted hybrid inflation model \cite{Khalil:2010cp}. As compared to this latter one, the smooth  hybrid inflation proposed in the present work generally predicts large tensor modes. 

Next, we discuss in brief  the implementation of the douplet-triplet solution to the well known issue of the color triplets $D_h, \bar{D}_{\bar{h}}$ embedded in the same representations $5$ and $\bar 5$ with the MSSM Higgs fields. 
The relevant superpotential terms are 
\be
W \supset \gamma \frac{\bar{h} \Phi^3 h}{m_{P}^{2}} + \delta \bar{h} h~.
\ee
After the symmetry breaking, these can be written in terms of the MSSM fields as follows
\be
W \supset \left( \delta - \frac{9 \gamma}{40\sqrt{15}} \frac{M^3}{m_{P}^{2}} \right) h_u h_d + \left(\delta + \frac{\gamma}{15\sqrt{15}} \frac{M^3}{m_{P}^{2}} \right) \bar{D}_{\bar{h}} D_h \supset \mu h_u h_d + M_{D_h} \bar{D}_{\bar{h}} D_h~.
\ee
\begin{figure}[t]
	\centering
	\begin{subfigure}[b]{0.495\textwidth}
		\centering
		\includegraphics[width=\textwidth]{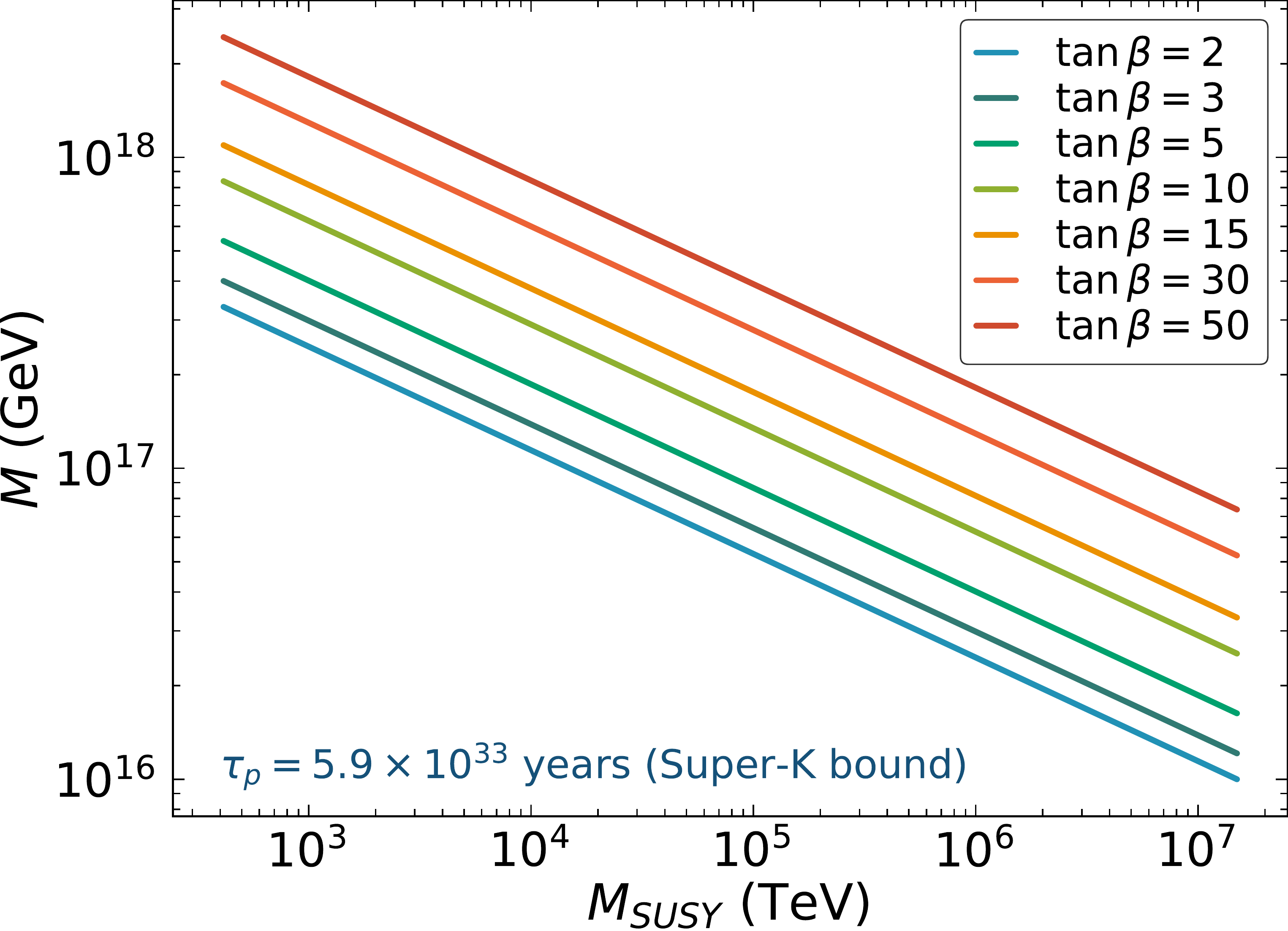}
		\caption{}
		\label{fig:superk_bound}
	\end{subfigure}
	\begin{subfigure}[b]{0.495\textwidth}
		\centering
		\includegraphics[width=\textwidth]{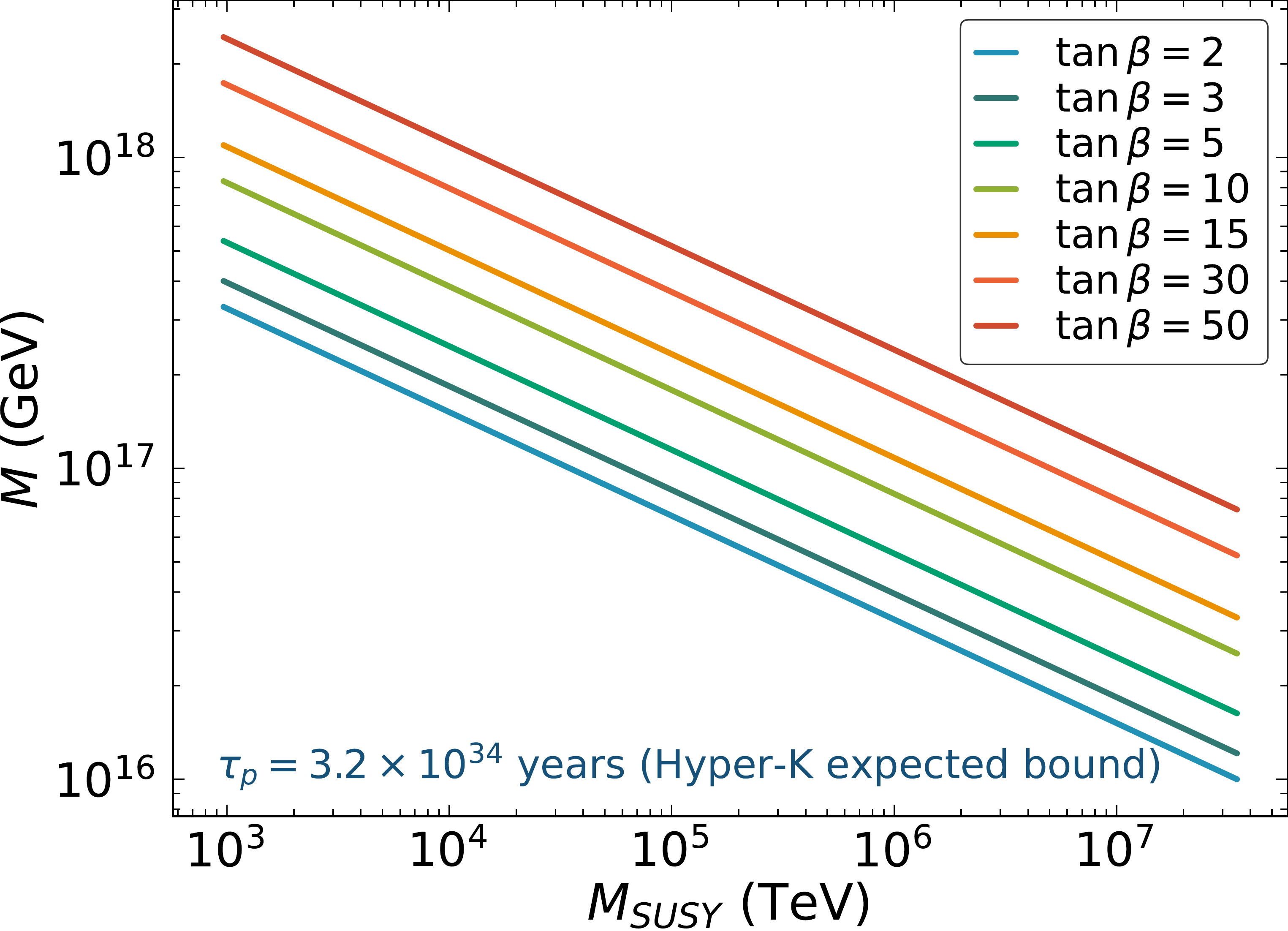}
		\caption{}
		\label{fig:hyperk_bound}
	\end{subfigure}
	\caption{$SU(5)$ gauge symmetry breaking scale $M$ as a function of SUSY breaking scale $M_{\text{SUSY}}$ for different values of $\tan \beta$, using Eq. \eqref{proton_lifetime}. The panel (a) is drawn for proton lifetime fixed at Super-Kamiokande bounds ($\tau_p = 5.9 \times 10^{33}$ years) while panel (b) is drawn for proton lifetime fixed at Hyper-Kamiokande future expected bounds ($\tau_p = 3.2 \times 10^{34}$ years).}
	\label{fig:proton_decay}
\end{figure}
We observe  that 
 the doublet-triplet splitting problem is effectuated by requiring fine-tuning of the involved parameters, 
such that
 $$\delta \sim \frac{9 \gamma}{40\sqrt{15}} \frac{M^3}{m_{P}^{2}}~.$$
Here $\mu$ is the MSSM $\mu$ parameter taken to be of the order of TeV scale while, $M_{D_h}$ is the color triplet Higgs mass parameter given by,
	\begin{equation}
	M_{D_h} \sim \frac{7 \, \gamma}{24\sqrt{15}} \frac{M^3}{m_{P}^{2}}.
	\end{equation}  
	The dominant contribution to proton decay amplitude comes from color-triplet Higgsinos. The decay rate from dimension 5 operator mediated by color-triplet Higgsinos typically dominates the decay rate from gauge boson mediated dimension 6 operators. The proton lifetime for the decay $p \rightarrow K^+ \bar{\nu}$ mediated by color-triplet Higgsinos is approximated by the following formula~\cite{Nagata:2013ive}:
	\begin{equation}
	\tau_p \simeq 4 \times 10^{35} \times \sin^4 2\beta \left( \frac{M_{\text{SUSY}}}{10^2 ~ \text{TeV}} \right)^2 \left( \frac{M_{D_h}}{10^{16} ~ \text{GeV}} \right)^2 \text{yrs}. \label{proton_lifetime}
	\end{equation}
	The proton lifetime depends on Higgino mass as well as the SUSY breaking scale $M_{SUSY}$. The experimental bounds on the proton lifetime can be satisfied at High scale SUSY.

Figure \ref{fig:proton_decay} shows $SU(5)$ gauge symmetry breaking scale $M$ as a function of SUSY breaking scale $M_{SUSY}$. The curves represent different values of $\tan \beta$. The left panel is drawn for proton lifetime fixed at current Super-Kamiokande bounds~\cite{Super-Kamiokande:2016exg} whilst the right panel is drawn for the future expected proton lifetime sensitivity limit\footnote{We emphasize that here we consider the Hyper-Kamiokande future expected limits in the case of non-observation of proton decay signals in order to show the status of proton decay in our model from future prospective.} of Hyper-Kamiokande experiment~\cite{Hyper-Kamiokande:2018ofw}. If the split and high scale SUSY is implemented, proton decay rate is within the acceptable bounds. The range of $M$ obtained in our numerical results in section \ref{sec5} is consistent with the experimental bounds on proton lifetime.

\section{\large{\bf Smooth hybrid $SU(5) \times U(1)_{\chi}$ inflation}}\label{sec3}
We will compute the effective scalar potential considering contributions from the F- and D-term sectors. The superpotential terms relevant for inflation are
\bea%
W &\supset& S\left( \mu^2 + \frac{Tr(\Phi^3)}{m_{P}}
\right) + \gamma \frac{\bar{h} \Phi^3 h}{m_{P}^{2}} + \delta \bar{h} h 
+ \sigma_{\chi} S \left( \chi \bar{\chi} - \mu_{\chi}^2 \right)+\zeta_{ij} {\chi} \nu_{i}^{c} \nu_{j}^{c}~. \label{superpotential_inflation}
\eea %
In component form, the above superpotential is expanded as follows,%
\bea%
W \supset S\left( \mu^2 + \frac{1}{4 m_P} d_{ijk}\phi_{i}\phi_{j}
\phi_{k}\right) &+& \delta \bar{h}_{a}h_{a} + \gamma \frac{\bar{h}_{a} h_{d}}{m_{P}^{2}} T^{i}_{ab} T^{j}_{bc} T^{k}_{cd} \, \phi_{i} \phi_{j} \phi_{k} \nonumber \\
&+& \sigma_{\chi} S \left( \chi \bar{\chi} - \mu_{\chi}^2 \right)+\zeta_{ij} {\chi} \nu_{i}^{c}\nu_{j}^{c},
\label{superpot-shift}%
\eea %
where $\Phi = \phi_i T^i$ with Tr$[T_i T_j] = \frac{1}{2}\delta_{ij}$ and $d_{ijk} = 2$Tr$[T_i\{T_j,T_k\}]$ in the $SU(5)$ adjoint basis. The $F$-term scalar potential obtained from the above superpotential is given by %
\begin{eqnarray}%
V_F &=&  \left| \; \mu^2 + \frac{1}{4 m_P}d_{ijk}\phi_{i}\phi_{j}
\phi_{k} + \sigma_{\chi} \left( \chi \bar{\chi} - \mu_{\chi}^2 \right) \; \right|^{2} \nonumber \\
&+&\sum_{i}\left| \; \frac{3}{4 m_P}d_{ijk} S \phi_{j} \phi_{k} + 3 \, \gamma \frac{\bar{h}_{a} h_{d}}{m_{P}^{2}} T^{i}_{ab} T^{j}_{bc} T^{k}_{cd} \, \phi_{j} \phi_{k}  \; \right|^{2} \nonumber \\ 
&+& \sum_{d}\left|\delta \bar{h_{d}} + \gamma \frac{\bar{h}_{a}}{m_{P}^{2}} T^{i}_{ab} T^{j}_{bc} T^{k}_{cd} \, \phi_{i} \phi_{j} \phi_{k} \right|^{2}  + \sum_{d}\left|\delta
h_{d} + \gamma \frac{h_{a}}{m_{P}^{2}} T^{i}_{ab} T^{j}_{bc} T^{k}_{cd} \, \phi_{i} \phi_{j} \phi_{k} \right|^{2} \nonumber\\
&+&  \left| \, \sigma_{\chi} S \bar{\chi} + \zeta_{ij}  \nu_{i}^{c}\nu_{j}^{c}\, \right|^2 + \left| \, \sigma_{\chi} S \chi \, \right|^2 + \left|2 \zeta_{ij} {\chi} \nu_{i}^{c}\right|^2 ,
\label{scalarpot-shift}
\end{eqnarray}%
where the scalar components of the superfields are denoted by the same symbols as the corresponding superfields. The VEV's of the fields at the global SUSY minimum of the above potential are given by,
\begin{gather}
S^0 = h_{a}^0  =  \bar{h_{a}^0} = \nu_{i}^{c\,0}=0, \nonumber \\ \;\; \chi^0 = \bar{\chi}^0 = \mu_{\chi}, \;\; Tr[(\Phi^0)^3] = d_{ijk}\phi_{i}^0\phi_{j}^0\phi_{k}^0 = -M^3/\sqrt{15} ,
\label{gmin}
\end{gather} 
where 
\bea
M = \left[4 \sqrt{15} \, m_P \, \mu^2  \right]^{1/3}. 
\label{M_definition}
\eea
\begin{figure}[t]
	\centering \includegraphics[width=9cm]{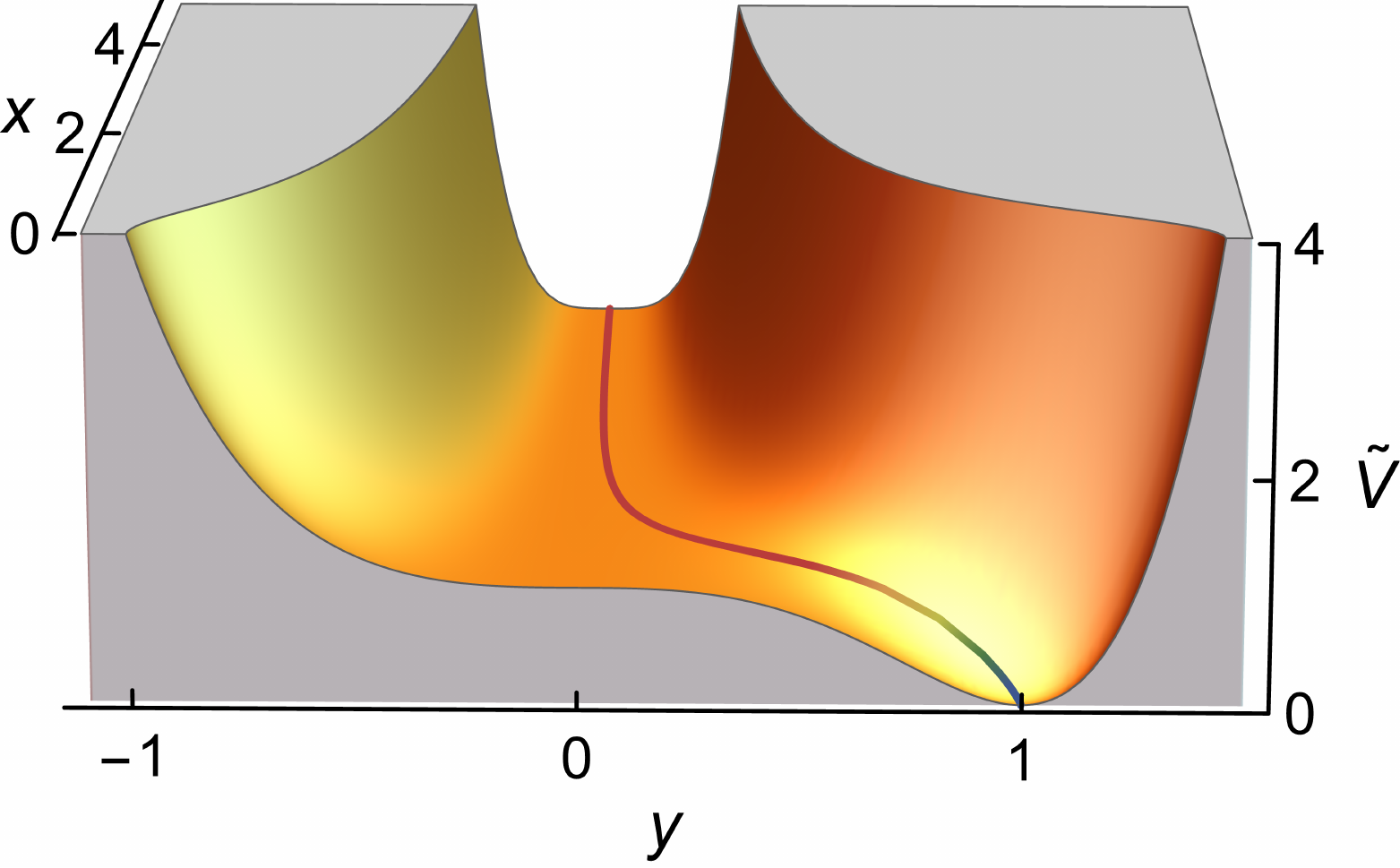}
	\caption{The tree-level, global scalar potential 
		$\tilde{V}=V/\mu^4$ of smooth $SU(5) \times U(1)_{\chi}$ hybrid inflation, with $\sigma_{\chi} \simeq 10^{-4
		}$, $\mu_{\chi} \simeq \mu \simeq 10^{14}$ GeV and $M \simeq M_{\text{GUT}} = 2 \times 10^{16}$ GeV.}
	\label{fig1}
\end{figure}
The superscript `0' denotes the field value at its global minimum. The superfield pair $\chi, \bar{\chi}$ which is $SU(5)$ singlet acquires an intermediate scale VEV such that $U(1)_{\chi}$ breaks to  $Z_2$, the matter parity. This symmetry ensures the existence of a  lightest supersymmetric particle (LSP) which could play the r\^ole of  a cold dark matter component.  Further, as discussed in~\cite{apal:2019}, this $Z_2$  symmetry yields topologically stable cosmic strings. Using $SU(5)$ symmetry transformation the VEV matrix $\Phi^0 = \phi_i^0 T^i$ can be aligned in the $24$-direction,
\be
\Phi_{24}^0 = \frac{\phi_{24}^0}{\sqrt{15}} \left( 1, 1, 1, - 3/2, - 3/2 \right).
\ee
This implies that $\phi_i^0 = 0, \, \forall \, i \neq 24$ and $\phi_{24}^0 = M $, where $d_{24\,24\,24} = -1/\sqrt{15}$ and $\phi_i^{0*} = \phi_i^0$ have been assumed. Thus the $SU(5)$ gauge symmetry is broken down to Standard Model gauge group $G_{\text{SM}}$ by the non-vanishing VEV of $\phi_{24}^0$ which is a singlet under $G_{\text{SM}}$.

The $D$-term scalar potential,
\bea %
V_D &=& \frac{g_{5}^2}{2} \sum_{i} \left( f^{ijk} \phi_j \phi_{k}^{\dagger}  + T^i \left( \left| h_a \right|^2  - \left| \bar{h}_a \right|^2  \right) \right)^2 \nonumber \\ 
&+& \frac{g_{\chi}^2}{2} \left( q_{\chi} \left| \chi \right|^2 + q_{\bar{\chi}} \left| \bar{\chi} \right|^2 + \left( q_{\bar{\chi}} + q_{\chi} \right) \varsigma \right)^2,
\eea %
also vanishes for this choice of the VEV (since $f^{i, 24, 24} = 0$) and for $\vert \bar{h}_a \vert = \vert h_a \vert$, $\vert \bar{\chi} \vert = \vert \chi \vert$. Here the symbol $\varsigma$ stands for the Fayet-Iliopoulos coupling constant.

The scalar potential in Eq.~(\ref{scalarpot-shift}) can be written in terms of the dimensionless variables%
\be %
x = \frac{|S|}{M}~, ~~~~~~~~~~~~ 
y = \frac{\phi_{24}}{M}~ ,~~~~~~~~~~~ 
z = \frac{\chi}{M}~,
\ee %
as follows, 
\be %
V = \mu^4 \left( \left(1-y^3 + \frac{\sigma_{\chi}}{\mu^2} \left( M^2 z^2 - \mu_{\chi}^2  \right) \right)^2 +  9 x^2 y^4 \right) + 2 \sigma_{\chi}^2 M^4 z^2  x^2 ~.%
\label{VF0}
\ee %

Because the potential \eqref{scalarpot-shift} is a complicated function of the fields $\phi_{24}$, $S$, $\chi$, $\bar{\chi}$, in order to make the computation of the extrema tractable, we have assumed $\chi = \bar{\chi} = \mu_{\chi}$, in accordance with their VEVs at the global SUSY minimum. We have checked with specific numerical examples that this simplification preserves the main features of the inflationary analysis in the parameter region of our interest. Indeed, this is due to the fact that in the present model, the $U(1)_{\chi}$ symmetry is broken after the $SU(5)$ symmetry so that the breaking scale $M$ of $SU(5)$ is always higher than the breaking scale $\mu_{\chi}$ of $U(1)_{\chi}$ which does not affect the inflation trajectory. Therefore, the potential can safely be stabilized at $z = \mu_{\chi}/M$. It can readily be checked that for $\mu_{\chi} \leq M$, the inflationary trajectories, in particular, remain unaltered in the $y-x$ plane. For $\mu_{\chi} > M$ however, the inflationary trajectory is affected and the potential is lifted. 
\noindent 
The potential is displayed in Fig. \eqref{fig1} which shows a valley of minimum given  by
\be
y=\frac{1}{{2}^{1/3}}\left(({\sqrt{32 x^6+1}+1})^{1/3}-({\sqrt{32 x^6+1}-1})^{1/3}\right)
\ee 
which in the large $x$ limit can be approximated as
$y\approx \frac{1}{6 \, x^2}.$

This valley of local minimum is not flat and possess a slope to drive inflaton towards SUSY vacuum.
Here we assume special initial conditions for inflation to occur in the valley~\footnote{See however~\cite{Tetradis:1997kp}  references therein for a  detailed discussion of the fine-tuning of initial conditions in various models of SUSY hybrid inflation.}. Implementing the large $x$ behaviour ($x \gg 1$) during inflation  the global SUSY potential~(\ref{VF0}) takes the form,
\be %
V \simeq \mu^4 \left( 1 - \frac{1}{432 \, x^6 } \right) + 2\, \mu_{\chi}^2 \sigma_{\chi}^2 M^2 x^2  .%
\ee %

\section{\large{\bf Reheating with non-thermal leptogenesis}}\label{sec4}

As soon as  the  inflation epoch ends, the vacuum energy is transfered to the energies of  coherent oscillations of the  inflaton $S$ and the scalar field $\theta=(\chi+\bar{\chi})/\sqrt{2}$ whose  decays give rise to the radiation in the universe. Reheating with leptogenesis, requires the presence of lepton-number violating interactions with the inflaton which  decays  into right-handed neutrinos. The latter are involved in the following superpotential terms
\begin{equation}
W \supset \sigma_{\chi} S \chi \bar{\chi} + \zeta_{ij} \chi \nu_{i}^c \nu_{j}^c +\overline{\zeta_{ij}}\frac{ \chi \chi }{\Lambda}\nu_{i}^c \nu_{j}^c ,\label{Infnu}
\end{equation}
where $\overline{\zeta_{ij}}$ is a coupling constant and $\Lambda$ represents a high cut-off scale (in a string model this could be identified with  the compactification scale). Heavy Majorana masses for the right-handed neutrinos are provided by the two last terms of~(\ref{Infnu})
\begin{equation}
	M_{\nu^c_{ij}}=\zeta_{ij} \langle \chi \rangle +\overline{\zeta_{ij}}\frac{\langle \chi \rangle \langle \chi \rangle }{\Lambda}~\cdot 
\end{equation}
Also, Dirac neutrino masses of the order of the electroweak scale  are obtained from the  tree-level superpotential term  $y_{ij}^{(\nu)}\,\nu_{i}^c\,\bar{f}_j\,h\to m_{ij}^{\nu_D}\nu\nu^c $   given in~(\ref{superpotential}).  Thus, the neutrino sector is
\be\label{Infnu2}
W\supset  {m_{\nu_D}}_{ij}\nu_i\nu_j^c+ M_{\nu^c_{ij}}\nu_i^c\nu_j^c.
\ee  
Note that the last term in~(\ref{Infnu2}) violates lepton number by two units, $\Delta L=2$. The small neutrino masses supported by neutrino oscillation experiments, are obtained by integrating out the heavy right-handed neutrinos and read as
\begin{equation}
{m_{\nu_D}}_{\alpha\beta}=-\sum_{i}{y^{(\nu)}}_{i\alpha}{y^{(\nu)}}_{i\beta}\frac{v_{u}^2}{M_i}~\cdot \label{mneu1}
\end{equation}
The neutrino mass matrix ${m_{\nu_D}}_{\alpha\beta}$ can be diagonalized by a unitary matrix $U_{\alpha i}$ as ${m_{\nu_D}}_{\alpha\beta} =  U_{\alpha i} U_{\beta i} m_{\nu_D}$, where $m_{\nu_D}$ is a diagonal
mass matrix $m_{\nu_D} = {\rm diag}(m_{\nu_{1}}, m_{\nu_{2}}, m_{\nu_{3}})$ and $M_{i}$ represent the eigenvalue of mass matrix  $M_{\nu^c_{ij}}$.

Another important implication of $\Delta L=2$ term is that lepton asymmetry is generated (inducing also baryon asymmetry~\cite{Fukugita:1986hr,Flanz:1994yx})
through right-handed neutrino decays, due to one-loop self-energy (CP-violating) diagrams. There  are contributions to two different decay channels  and the ratio of the lepton number density
to the entropy density  in the limit $T_r < M_{1}\leq m_{\text{inf}} /2 \leq M_{2,3}$ is defined as
\begin{equation}
\frac{n_{L}}{s}\sim \frac{3}{2}\frac{T_{r}}{m_{\text{inf}}}\epsilon_{cp}~,
\end{equation}
where $\epsilon_{cp}$ is the CP asymmetry factor and is generated from the out of equilibrium decay of lightest right-handed neutrino and is given by \cite{Hamaguchi:2002vc},
\begin{equation}
\epsilon_{cp}=-\frac{3}{8\pi}\frac{1}{\left(y^{(\nu)}y^{{(\nu)}\dagger}\right)_{11}}\sum_{i=2,3}Im\left[\left(y^{(\nu)}y^{{(\nu)}\dagger}\right)_{1i}\right]^2\frac{M_1}{M_i},
\end{equation}
and $T_{r}$ is reheating temperature which can be as estimated as
\begin{eqnarray}
T_r \simeq \sqrt[4]{\frac{90}{\pi^2 g_{\star}}} \sqrt{\Gamma \, m_P}~,
\label{reheat}
\end{eqnarray} 
where $g_{\star}$ is $228.75$ for MSSM. The $\Gamma$ is the decay width for the inflaton decay into right-handed neutrinos and is given by \cite{Hamaguchi:2002vc}
\begin{equation}
\Gamma \left({\rm inf} \rightarrow \nu_{i}^c \nu_{j}^c \right) = \frac{1}{8 \pi}\left(\frac{M_1}{\mu_{\chi}}\right)^2 \, m_{\text{inf}}  \left(  1 - \frac{4 M_{1}^2}{m_{\text{inf}}^2} \right)^{1/2},
\end{equation}
with the inflaton mass given by
\begin{equation}
m_{\text{inf}} = \sqrt{\frac{9 \mu^4}{M^2} + 2 \mu_{\chi}^2 \sigma_{\chi}^2}
\, \,.
\end{equation}
Assuming a normal hierarchical pattern of light neutrino masses, the CP asymmetry factor, $\epsilon_{cp}$, becomes 
\begin{equation}
\epsilon_{cp} = \frac{3}{8\pi}\frac{M_1 m_{\nu_{3}}}{v_{u}^2}\delta_{\rm eff}, 
\end{equation}
where $m_{\nu_3}$ is the mass of the heaviest light neutrino, $v_{u}=\langle H_u \rangle $ is the VEV of the up-type electroweak Higgs and $\delta_{\rm eff}$ is the CP-violating phase. The experimental value of lepton asymmetry is estimated as \cite{Planck:2018vyg},
\begin{eqnarray}
\mid n_L/s\mid\approx\left(2.67-3.02\right)\times 10^{-10}.
\end{eqnarray}
 In the numerical estimates discussed below we take $m_{\nu_3} = 0.05$ eV, $|\delta_{\rm eff}|=1$ and $v_u = 174$ GeV, while assuming large $\tan \beta $.
The non-thermal production of lepton asymmetry, $n_{L}/s$, is given by the following expression  
\begin{equation}
\frac{n_L}{s} \lesssim 3 \times 10^{-10} \frac{T_r}{m_{\text{inf}}}\left(\frac{M_1}{10^6 \text{ GeV}}\right)\left(\frac{m_{\nu_3}}{0.05 \text{ eV}}\right) \label{nls},
\end{equation}
with $M_{1} \gg T_r $. Using the experimental value of $n_L/s\approx 2.5\times 10^{-10}$ with Eq. \eqref{reheat} and \eqref{nls}, we obtain the following lower bound on $T_r$,
\begin{equation}
T_r \gtrsim 1.9 \times 10^7 \text{ GeV}  \left(\frac{m_{\text{inf}}}{10^{11}\text{ GeV}}\right)^{3/4} \left(\frac{10^{16} \, \text{GeV}}{M_1}\right)^{1/2}\left(\frac{m_{\nu_3}}{0.05 \text{ eV}}\right)^{1/2} \label{lepto}.
\end{equation}
 A successful baryogenesis is usually generated through the sphaleron processe where an initial lepton asymmetry, $n_L/s$, is partially converted into a baryon asymmetry $n_{B}/s=-0.35n_L/s$  \cite{Khlebnikov:1988sr,Harvey:1990qw}. Eq. \eqref{lepto} is used in our numerical analysis to calculate inflationary predictions which are consistent with leptogenesis and baryogenesis.
\section{\large{\bf Minimal K\"ahler potential}}\label{sec5}

In this section we will include  SUGRA corrections in the effective potential and consider their implications in 
the inflationary parameters.
We first start with the minimal canonical K\"ahler potential,
\be \label{Ktree}
K = \vert S \vert^2 + Tr \vert \Phi \vert^2
+ \vert h \vert^2 + \vert \bar{h}\vert^2 + \vert \chi \vert^2 + \vert \bar{\chi} \vert^2 .
\ee
The F-term SUGRA scalar potential is given by 
\begin{equation}
	V_{\text{SUGRA}}=e^{K/m_P^{2}}\left(
	K_{i\bar{j}}^{-1}D_{z_{i}}WD_{z^{*}_j}W^{*}-3 m_P^{-2}\left| W\right| ^{2}\right),
	\label{VF}
\end{equation}
with $z_{i}$ being the bosonic components of the superfields $z%
_{i}\in \{S,\Phi,h,\bar{h}, \chi, \bar{\chi} ,\cdots\}$, and we have defined
\be
D_{z_{i}}W \equiv \frac{\partial W}{\partial z_{i}}+m_P^{-2}\frac{%
	\partial K}{\partial z_{i}}W , \,\,\,
K_{i\bar{j}} \equiv \frac{\partial ^{2}K}{\partial z_{i}\partial z_{j}^{*}},
\ee
and $D_{z_{i}^{*}}W^{*}=\left( D_{z_{i}}W\right)^{*}.$
The SUGRA scalar potential during inflation becomes
\begin{eqnarray}
	V_{\text{SUGRA}} &=& \mu^4 \, \left[ 1 - \frac{1}{432 \, x^6} + 2\, \left( \frac{\mu_{\chi}}{m_P} \right)^2  + 2\, \left( \frac{\mu_{\chi}}{m_P} \right)^4 + \frac{40}{3} \sigma_{\chi}^2 \left( \frac{\mu_{\chi}}{m_P} \right)^2 \right. \nonumber \\ \nonumber
	&+& \left.   2\, x^2 \left(\frac{M}{m_{P}}\right)^2 \left(\frac{\mu_{\chi}}{m_{P}}\right)^2 + \frac{x^4}{2} \left(\frac{M}{m_{P}}\right)^4 \right] + 2 \, M^4 x^4 \left[ \left(\frac{\mu_{\chi}}{m_{P}}\right)^2 + 2 \left(\frac{\mu_{\chi}}{m_{P}}\right)^4\right] \sigma_{\chi}^2  \\ \nonumber
	&+& 4\, \mu^2 \sigma_{\chi} \, \left[  x^4 \, \mu_{\chi}^2 \left(\frac{M}{m_{P}}\right)^4 +  x^2 M^2 \left( \left(\frac{\mu_{\chi}}{m_{P}}\right)^2 + 2 \, \left(\frac{\mu_{\chi}}{m_{P}}\right)^4  \right)  \right] \\
	&+& 2 \,  \mu_{\chi}^2 M^2 x^2 \left[ 1  + 2 \, \left(\frac{\mu_{\chi}}{m_{P}}\right)^2  \right] \sigma_{\chi}^2 .
\end{eqnarray}

The inflationary slow roll parameters are given by,
\bea
\epsilon = \frac{1}{4}\left( \frac{m_P}{M}\right)^2
\left( \frac{V'}{V}\right)^2, \,\,\,
\eta = \frac{1}{2}\left( \frac{m_P}{M}\right)^2
\left( \frac{V''}{V} \right), \,\,\,
\xi^2 = \frac{1}{4}\left( \frac{m_P}{M}\right)^4
\left( \frac{V' V'''}{V^2}\right).
\eea
Here, the derivatives are with respect to $x=|S|/M$, whereas the canonically normalized field $\sigma \equiv \sqrt{2}|S|$. In the slow-roll (leading order) approximation, the tensor-to-scalar ratio $r$, the scalar spectral index $n_s$, and the running of the scalar spectral index $dn_s / d \ln k$ are given by
\bea
r &\simeq& 16\,\epsilon,  \\
n_s &\simeq& 1+2\,\eta-6\,\epsilon,  \\
\frac{d n_s}{d\ln k} &\simeq& 16\,\epsilon\,\eta
-24\,\epsilon^2 - 2\,\xi^2 .
\eea
\begin{figure}[t]
	\centering \includegraphics[width=8.10cm]{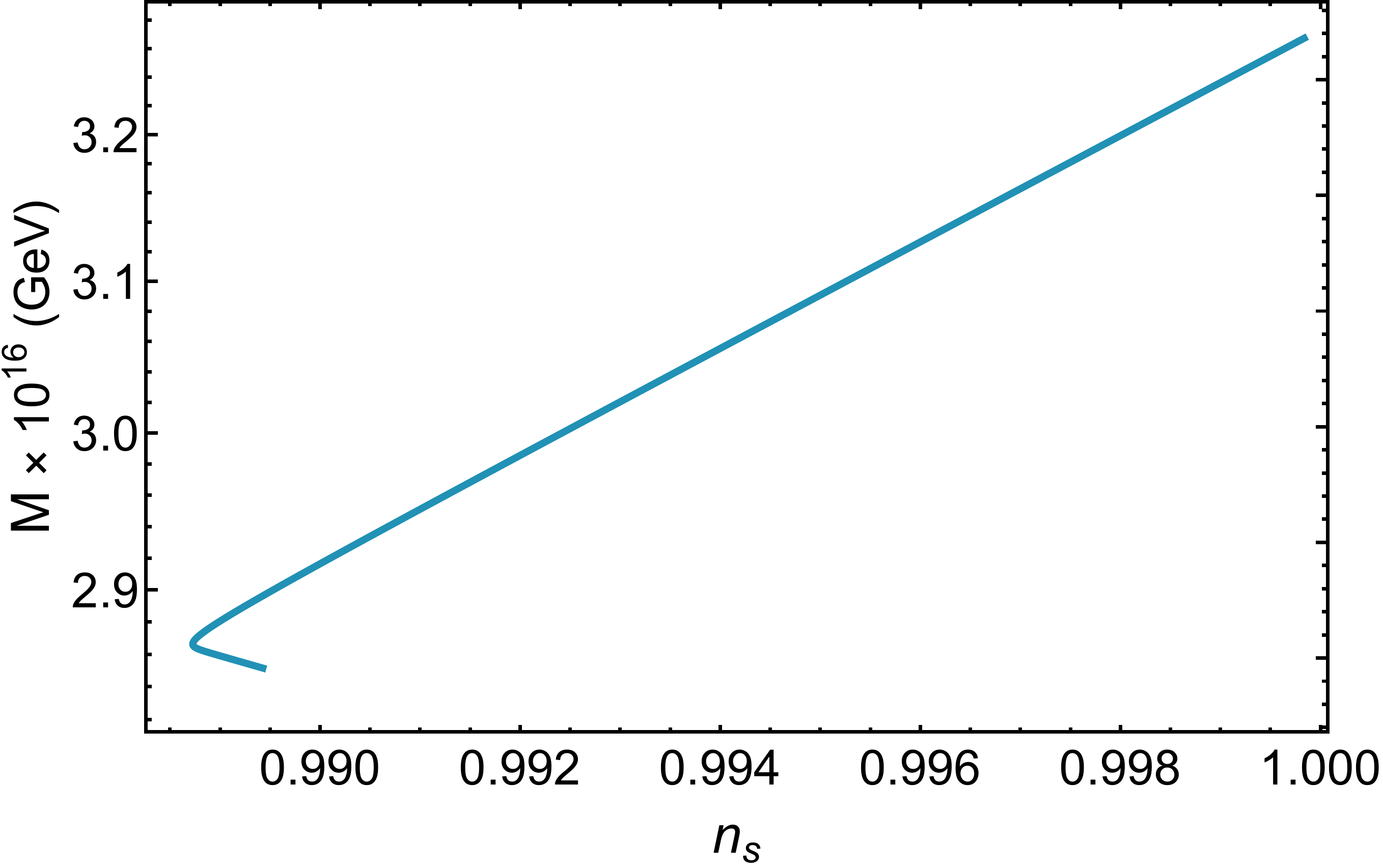}
	\centering \includegraphics[width=7.90cm]{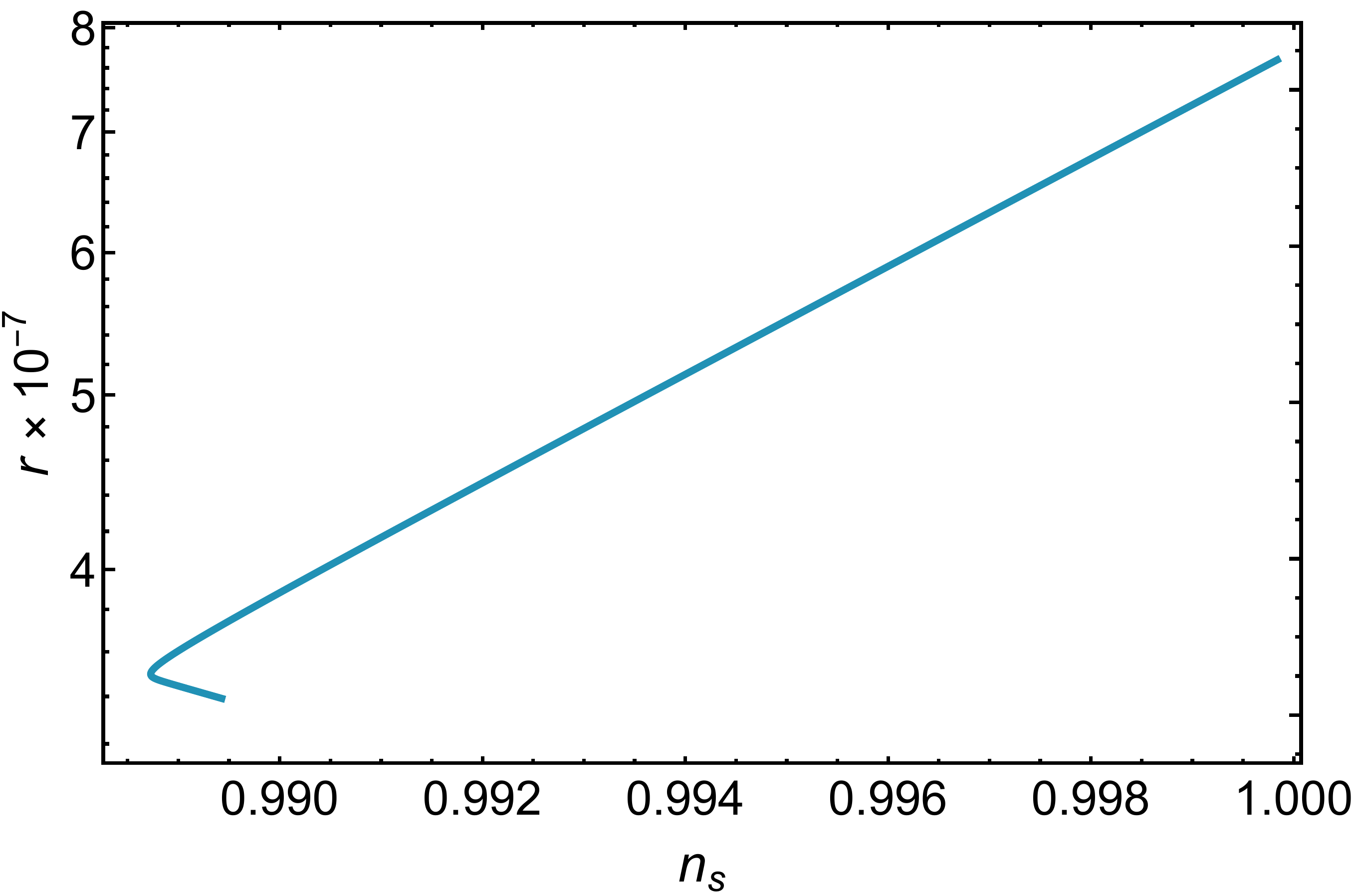}
	\centering \includegraphics[width=7.80cm]{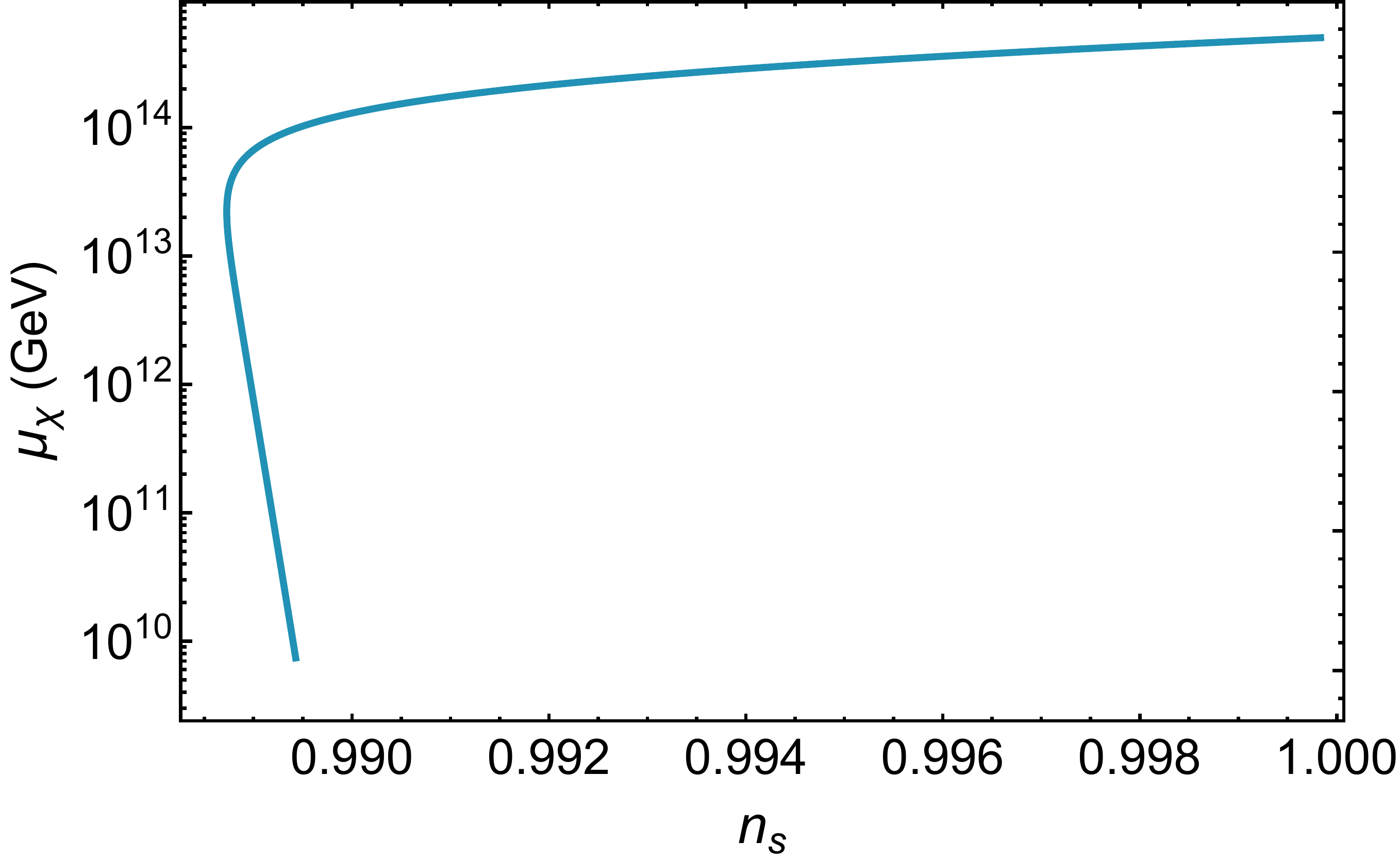}
	\centering \includegraphics[width=8.20cm]{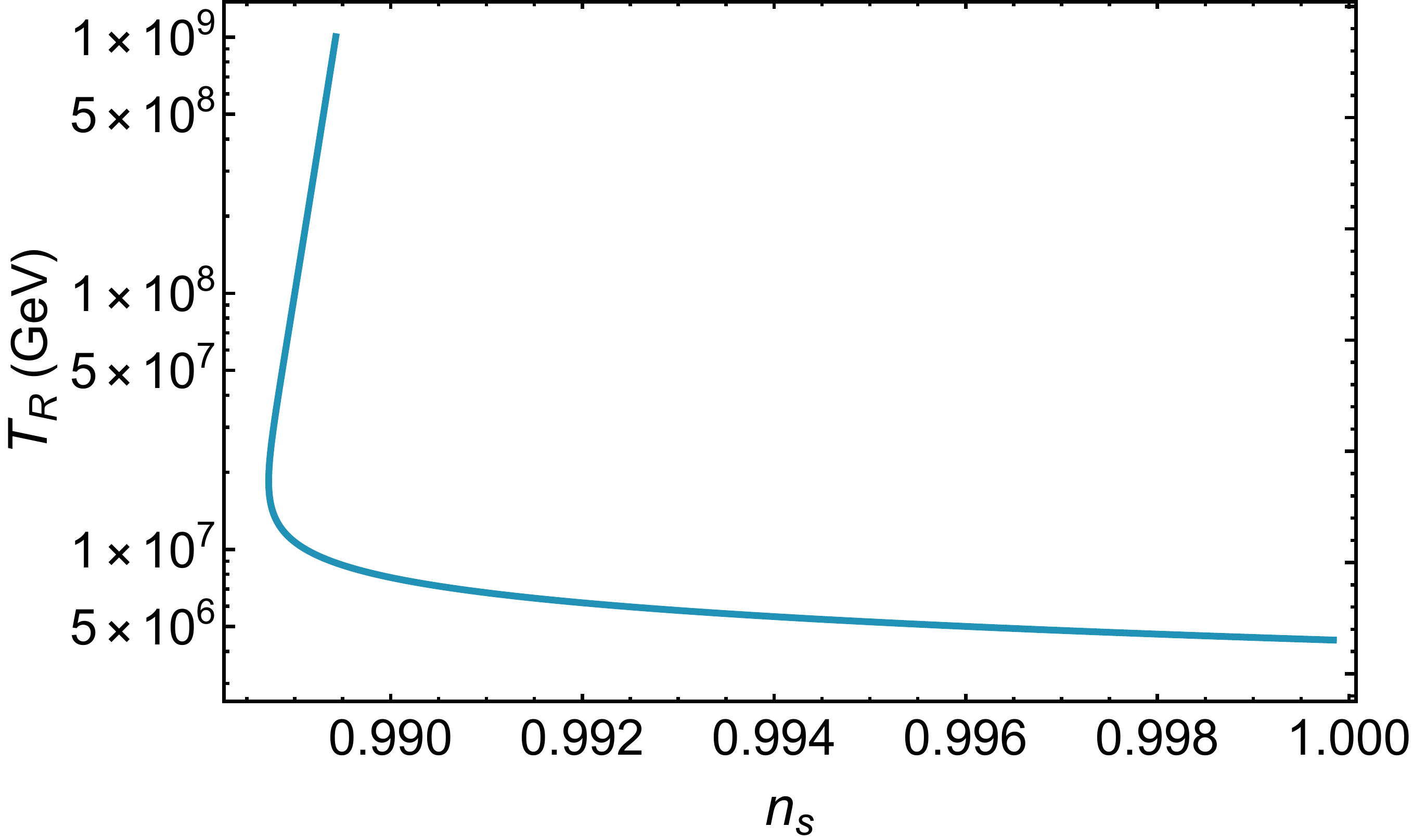}
	\caption{The scalar spectral index $n_s$ vs the $SU(5)$ symmetry breaking scale $M$, the tensor-to-scalar ratio $r$, the $U(1)_{\chi}$ symmetry breaking scale $\mu_{\chi}$ and the reheat temperature $T_r$ for minimal K\"ahler potential without the soft mass terms.}
	\label{minimal_nosoft1}
\end{figure}
The last $N_0$ number of e-folds before the end of inflation is,
\bea
N_0 = 2\left( \frac{M}{m_P}\right) ^{2}\int_{x_e}^{x_{0}}\left( \frac{V}{%
	V'}\right) dx,
\label{efolds}
\eea
where $x_0$ is the field value at the pivot scale $k_0$, and
$x_e$ is the field value at the end of inflation, defined by $|\eta(x_e)| = 1$.  Assuming a standard thermal history, $N_0$ is related to $T_r$ as \cite{Garcia-Bellido:1996egv}
\begin{equation}\label{n0}
	N_0=54+\frac{1}{3}\ln\Big(\frac{T_r}{10^9\text{ GeV}}\Big)+\frac{2}{3}\ln\Big(\frac{V(x)^{1/4} }{10^{15}\text{ GeV}}\Big),
\end{equation}
where $T_r$ is given by Eq.\eqref{reheat}. The amplitude of the curvature perturbation is given by \cite{Liddle:1993fq}
\bea
A_{s}(k_0) = \frac{1}{24\,\pi^2}
\left. \left( \frac{V/m_P^4}{\epsilon}\right)\right|_{x = x_0},
\label{perturb}
\eea
where $A_{s}= 2.137 \times 10^{-9}$ is the Planck normalization at $k_0 = 0.05\, \rm{Mpc}^{-1}$. The one-loop radiative corrections are expected to have a negligible effect on the inflationary predictions; we can therefore ignore these contributions in our numerical calculations. Fig. \ref{minimal_nosoft1}, shows the results without soft SUSY mass terms. It can be seen that without the soft mass terms, it is not possible to obtain $n_s$ within Plank 2-$\sigma$ bounds. We resolve this problem by including the soft mass terms, whose effect on the inflationary predictions have been assumed negligible in previous studies \cite{Senoguz:2003zw}. With the inclusion of soft mass terms, the scalar spectral index $n_s$ is easily obtained within Planck's 2-$\sigma$ bounds. We consider gravity-mediated SUSY breaking scenario, where SUSY is broken in the hidden sector and is communicated gravitationally to the observable sector. Following \cite{Nilles:1983ge},  the soft potential is
\begin{figure}[t]
	\centering \includegraphics[width=7.90cm]{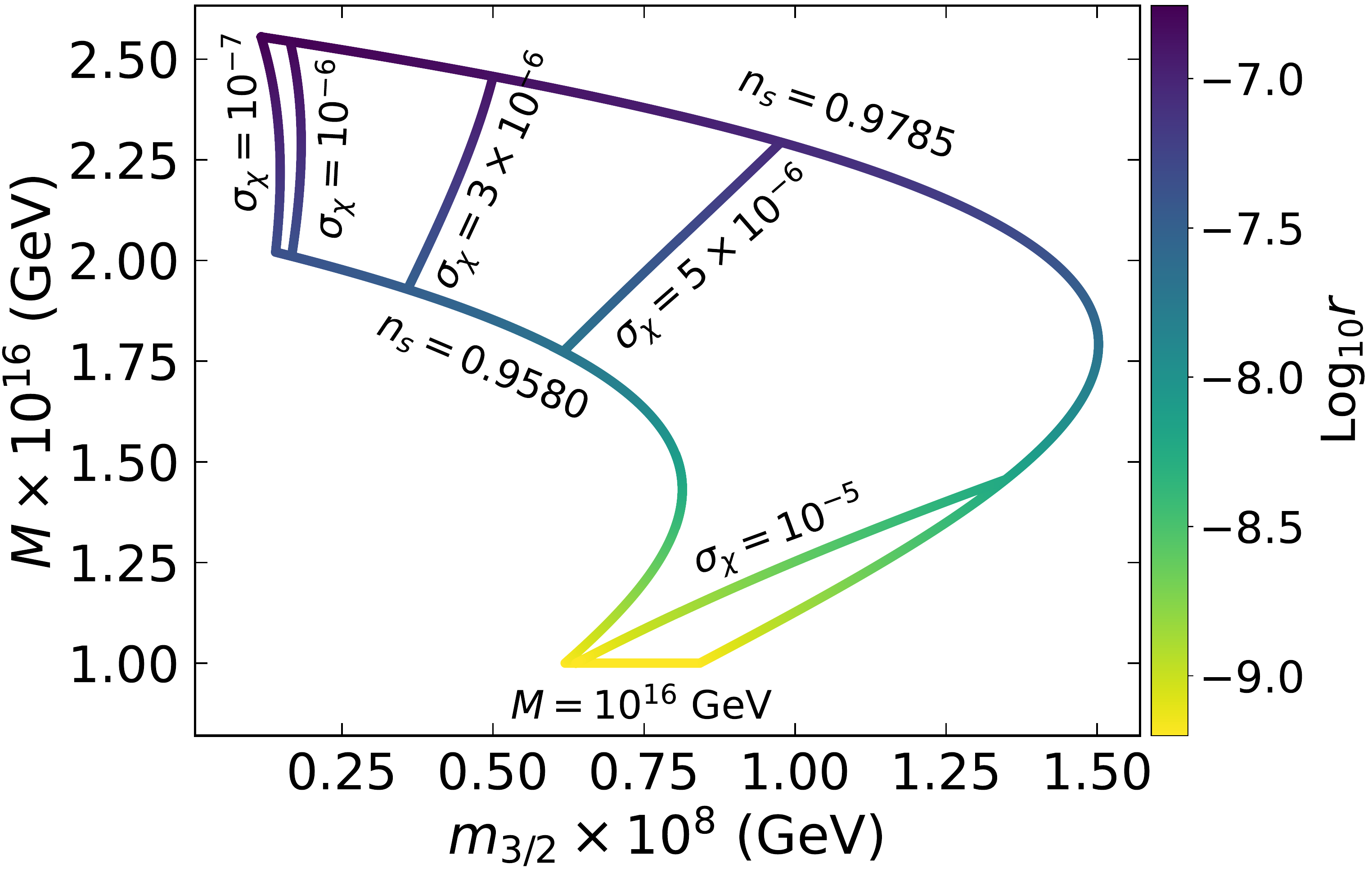}
	\centering \includegraphics[width=7.90cm]{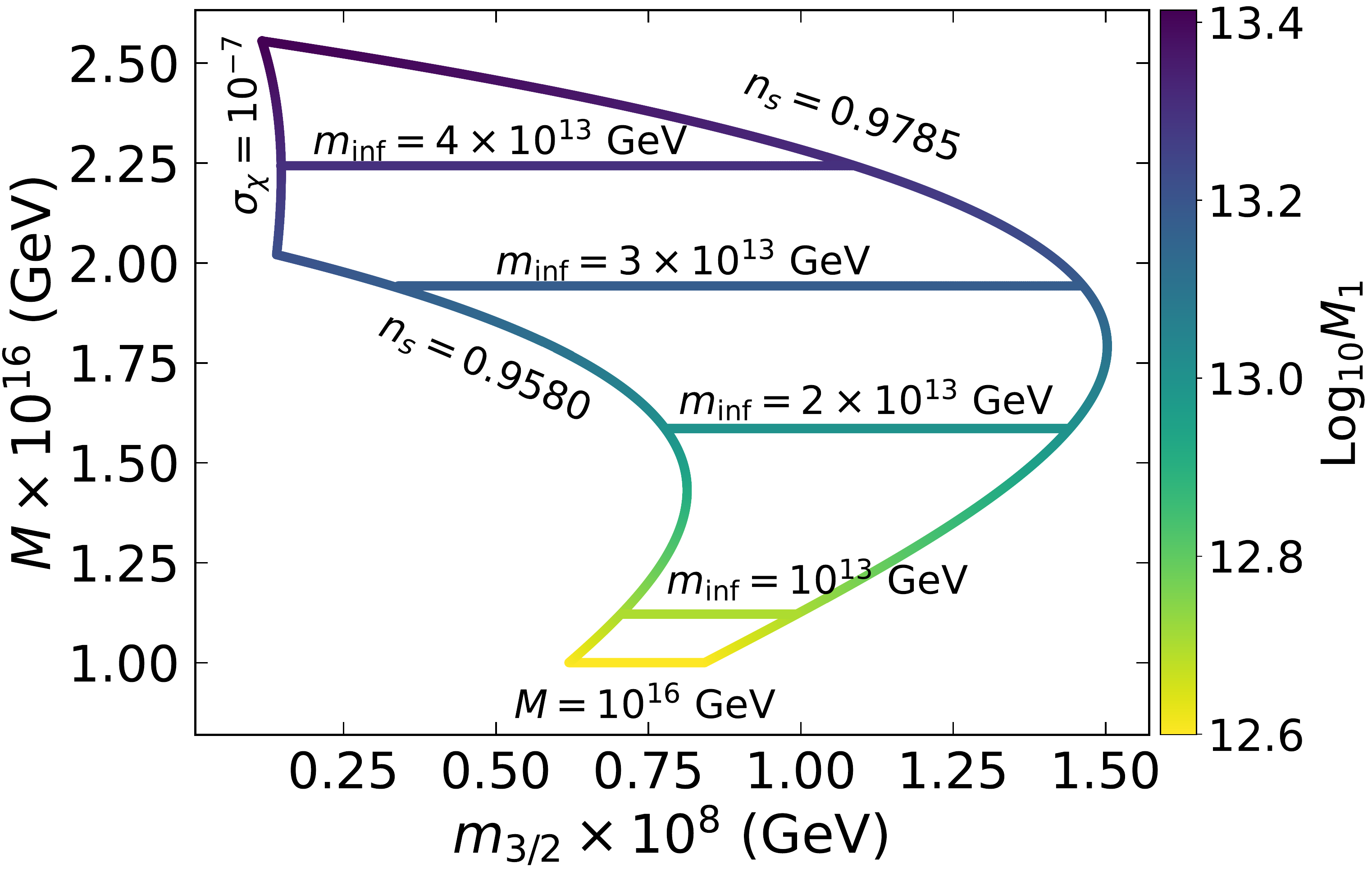}\\ \vspace*{5pt}
	\centering \includegraphics[width=7.80cm]{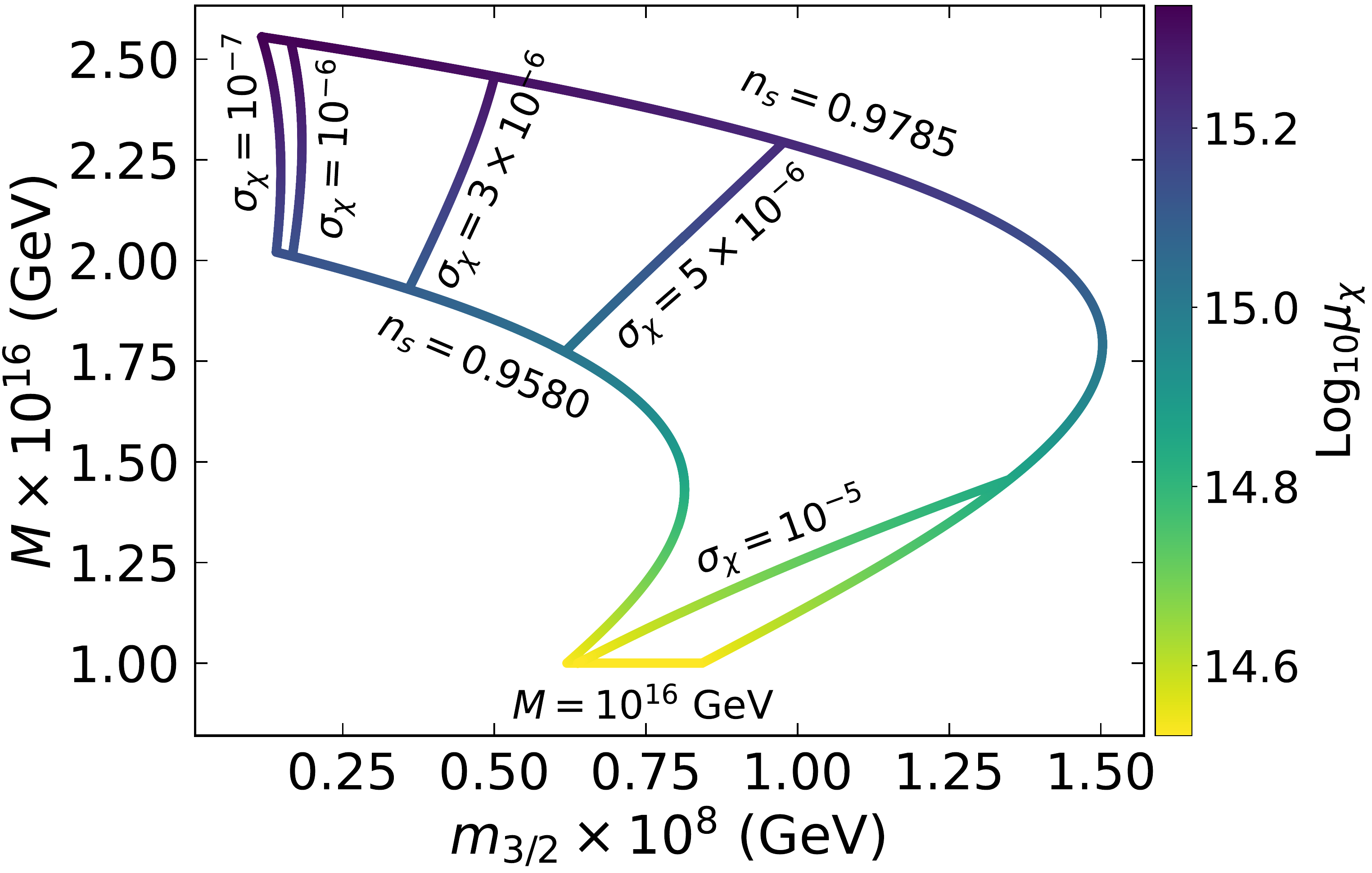}
	\caption{The variation of various parameters in the $SU(5)$ symmetry breaking scale M and gravitino mass $m_{3/2}$ plane. By including the contribution of soft mass terms, the scalar spectral index $n_s$ lies within Planck's 2-$\sigma$ bounds.}
	\label{minimal_witsoft1}
\end{figure}
\bea
V_{Soft} = M_{z_{i}}^2\vert z_{i} \vert^2+m_{3/2}\left\{z_{i}W_{i}+\left(A-3\right)W +h.c\right\},
\eea
where $z_{i}$ is observable sector field, $W_{i}=\frac{\partial W}{\partial z_{i}}$, $m_{3/2}$ is the gravitino mass and $A$ is the complex coefficient of the trilinear soft-
SUSY-breaking terms. The effective scalar potential during inflation then reads as,
\begin{eqnarray}
	V_{\text{eff}} = V_{\text{SUGRA}} + a \mu^2 M m_{3/2} x+ b \mu_{\chi}^2 M m_{3/2} \sigma_\chi x + 2M^2 M_{S}^2 x^2+\mu_\chi^2 M_{\chi}^2,
\end{eqnarray}
where $a=2\vert A-2 \vert \cos \left(\arg S + \arg \vert A-2 \vert\right)$ and $b=2 \cos \left(\arg S\right)$. For convenience, we have set $a = b$ and $M_{S} = M_{\chi} = m_{3/2}$. The soft mass terms, seem to play an important r\^ole in inflationary predictions. Fig. \ref{minimal_witsoft1} shows our numerical results with soft mass terms where the variation of various parameters is depicted in the $m_{3/2}-M$ plane. It is evident that, with soft terms, the prediction of scalar spectral index $n_s$ lies within Planck's latest bounds. In the slow-roll approximation, the amplitude of the power spectrum of scalar
curvature perturbation $A_{s}$ as given by Eq \ref{perturb} is,
\bea
A_{s}(k_{0}) = \frac{1}{6\,\pi^2}\left(\frac{M}{m_{p}}\right)^{2}
\left.\left( \frac{V^{3}/V^{\prime\;2}}{m_P^{4}}\right)\right|_{x = x_0},
\eea
which becomes
\begin{equation}
	\begin{aligned}
		A_{s}(k_0) &\simeq 1440 \left( \frac{m_P}{M} \right)^4 \Bigg( \frac{60 \, a \, m_P^3 \, m_{3/2}}{\sqrt{15} M^3} + \left( \frac{m_P}{M} \right)^2 \left(\frac{1}{72 x_0^7} + 2 x_0^3 \left( \frac{M}{m_P} \right)^4  + 4 x_0 \frac{\mu_{\chi}^2 M^2}{m_P^4}  \right)  \\
		&+  \frac{480 \, x_0 \, \sigma_{\chi} \, \mu_{\chi}^2 \, m_P}{\sqrt{15} M^3} +960 \, x_0 \, \mu_{\chi}^2 \sigma_{\chi}^2 \left( \frac{m_P}{M} \right)^4     \Bigg).
	\end{aligned}\label{analytic_as}
\end{equation}
Also, the spectral index $n_s$  to leading
order is given by
\begin{eqnarray}
	n_{s}&\simeq&  1 +\frac{(4\sqrt{15})^{4/3}m_{p}^{4/3}}{(6\pi)^{1/3}M^{16/3}}\left(\frac{V^{\prime\prime}}{A_{s}^{1/3}}\right). \label{eq_ns}  
\end{eqnarray}
It can be checked that the analytical expressions in \eqref{analytic_as} and \eqref{eq_ns} are consistent with the numerical results displayed in Fig. \ref{minimal_witsoft1}. We obtain a low reheat temperature $T_r \simeq 10^6$ GeV consistent with leptogensis and baryongensis. With low reheat temperature, the gravitino problem can be avoided for all ranges of gravitino mass.
For scalar spectral index $n_s$ within Planck's 2-$\sigma$ bounds, the $SU(5)$ breaking scale $M$ is obtained in the range $(1 - 2.5) \times 10^{16}$ GeV. This in turn requires $m_{3/2} \gtrsim 10^6$ TeV in order to avoid  $d = 5$ rapid proton decay, as shown in Fig. \ref{fig:proton_decay}. The gravitino mass turns out to be in the range $(8.4 \times 10^4 - 1.5 \times 10^5)$ TeV, which is less than $10^6$ TeV required to avoid rapid proton decay. The minimal case is therefore not consistent with experimental bounds on proton lifetime. The exact calculation for $d = 5$ proton lifetime, however, may reduce the SUSY breaking scale $M_{\text{SUSY}}$ and hence $m_{3/2}$, avoiding the rapid proton decay problem. Furthermore, the tensor to scalar ratio $r$ turns out to be very small $r \lesssim 10^{-7}$ and is beyond the current measuring limits of various experiments like Planck \cite{Planck:2018jri}, LiteBIRD \cite{Matsumura:2013aja}, PIXIE \cite{Kogut:2011xw} and CORE \cite{CORE:2016ymi}. In order to find large tensor modes consistent with proton decay and non-thermal leptogensis, we employ non-minimal K\"ahler potential as discussed in the next section.

\section{\large{\bf Non-minimal K\"ahler potential}}\label{sec6}

In  effective theory models, non-renormalizable contributions convey information of the `higher theory', and as such, are expected to play a significant r\^ole.
In this  section we employ a non-minimal K\"ahler potential  including NR terms up to sixth order  in the tree-level one \eqref{Ktree}. Then, it takes the form

\begin{eqnarray}\small
	\label{K}
	\begin{split} 
K &= \vert S \vert^2 + Tr \vert \Phi \vert^2
+ \vert h \vert^2 + \vert \bar{h}\vert^2 + \vert \chi \vert^2 + \vert \bar{\chi} \vert^2
\\
  &+\kappa_{S\Phi} \frac{\vert S\vert^2 \, Tr \vert \Phi \vert^2}{m_P^2}
  + \kappa_{S h} \frac{\vert S \vert^2 \vert h \vert^2}{m_P^2}
  + \kappa_{S \bar{H}} \frac{\vert S \vert^2 \vert \bar{H} \vert^2}{m_P^2}
   + \kappa_{S {\chi}} \frac{\vert S \vert^2 \vert \chi \vert^2}{m_P^2} + \kappa_{S \bar{\chi}} \frac{\vert S \vert^2 \vert \bar{\chi} \vert^2}{m_P^2} + \kappa_{H \Phi} \frac{\vert h \vert^2 \, Tr \vert \Phi \vert^2}{m_P^2}
   \\
   & + \kappa_{h \chi} \frac{\vert h \vert^2  \vert \chi \vert^2}{m_P^2} + \kappa_{h \bar{\chi}} \frac{\vert h \vert^2  \vert \bar{\chi} \vert^2}{m_P^2} + \kappa_{\bar{h} \Phi} \frac{\vert \bar{h} \vert^2 \, Tr \vert \Phi \vert^2}{m_P^2}
   + \kappa_{\bar{h} \chi} \frac{\vert \bar{h} \vert^2  \vert \chi \vert^2}{m_P^2} + \kappa_{\bar{h} \bar{\chi}} \frac{\vert \bar{h} \vert^2  \vert \bar{\chi} \vert^2}{m_P^2} + \kappa_{h \bar{h}} \frac{\vert h \vert^2 \vert \bar{h} \vert^2}{m_P^2}
\\
  &  + \kappa_{\chi \bar{\chi}} \frac{\vert \chi \vert^2 \vert \bar{\chi} \vert^2}{m_P^2} + \kappa_S \frac{\vert S\vert^4}{4 m_P^2}
   + \kappa_{\Phi} \frac{ (Tr \vert \Phi \vert^2)^2}{4 m_P^2}
   + \kappa_{H} \frac{ \vert h \vert^4}{4 m_P^2} + \kappa_{\bar{h}} \frac{ \vert \bar{h} \vert^4}{4 m_P^2}
   + \kappa_{\chi} \frac{ \vert \chi \vert^4}{4 m_P^2}
   + \kappa_{\bar{\chi}} \frac{ \vert \bar{\chi} \vert^4}{4 m_P^2}
   \\
  & + \kappa_{SS} \frac{\vert S\vert^6}{6 m_P^4} + \kappa_{\Phi \Phi} \frac{ (Tr \vert \Phi \vert^2)^3}{6 m_P^4}
   + \kappa_{h h} \frac{ \vert h \vert^6}{6 m_P^4}
   + \kappa_{\bar{h} \bar{h}} \frac{ \vert \bar{h} \vert^6}{6 m_P^4}
   + \kappa_{\chi \chi} \frac{ \vert \chi \vert^6}{6 m_P^4} + \kappa_{\bar{\chi} \bar{\chi}} \frac{ \vert \bar{\chi} \vert^6}{6 m_P^4}
   + \cdots.
	\end{split}
\end{eqnarray}
As $\Phi$ is an adjoint superfield, many other terms of the form,
\bea
f\left(\vert S \vert^2, \vert \Phi \vert^2,\,\frac{Tr(\Phi^3)}{m_P} + h.c., \cdots \right) \, ,
\eea
can appear in the K\"ahler potential. The effective contribution of all these terms is either suppressed or can be absorbed into other terms already present in the K\"ahler potential.
Therefore, the supergravity (SUGRA) scalar potential during inflation becomes

\begin{eqnarray}
	V_{\text{SUGRA}} &=& \mu^4 \, \left[ 1 - \frac{1}{432 \, x^6} + 2\,  \left( 1 - \kappa_{S \chi} \right) \left( \frac{\mu_{\chi}}{m_P}  \right)^2 \right. \nonumber \\
	&+& \left. 2\,  \left( 1 + \frac{1}{4} \kappa_{\chi} -2 \, \kappa_{S \chi} \left( 1 - \kappa_{S \chi} \right) \right) \left( \frac{\mu_{\chi}}{m_P}  \right)^4 \right. \nonumber \\
	&+& \left. \frac{40}{3} \sigma_{\chi}^2 \left( 1 - \kappa_{S {\chi}} \right) \left( \frac{\mu_{\chi}}{m_P}  \right)^2 - \kappa_S \, x^2 \left(\frac{M}{m_{P}}\right)^2 + \gamma_S \, \frac{x^4}{2} \left(\frac{M}{m_{P}}\right)^4  \right. \nonumber \\
	&+& \left. 2 \, x^2 \left( 1 - \kappa_{S} - 2 \kappa_{S {\chi}} \left( 1 - \kappa_{S} \right) + \kappa_{S {\chi}}^2 \right) \left( \frac{M}{m_P}  \right)^2 \left( \frac{\mu_{\chi}}{m_P}  \right)^2  \right]   \nonumber \\
	&+& 4\, \mu^2 \, \sigma_{\chi} \mu_{\chi}^2 \left[  \left( 1 - \kappa_{S \chi} \right) x^2 \left(\frac{M}{m_{P}}\right)^2 + \left(1 + \kappa_{S \chi} \left( \kappa_S  + \kappa_{S \chi} -2  \right)\right) x^4 \left(\frac{M}{m_{P}}\right)^4 \right] \nonumber \\
	&+& 8\, \mu^2 M^2 \sigma_{\chi}   \left( 1 - \frac{1}{4} \kappa_{\chi} + \kappa_{S \chi} \left( \kappa_{S \chi} - 1 + \frac{1}{2} \kappa_{\chi} \right) \right) x^2 \left(\frac{\mu_{\chi}}{m_P}\right)^4  \nonumber \\
	&+& 2 \, M^4 \sigma_{\chi}^2 x^4 \left[ \left( 1 - \kappa_{S \chi} \right)  \left(\frac{\mu_{\chi}}{m_{P}}\right)^2 + 2 \left( 1 + \kappa_{S \chi} \left( \kappa_{S {\chi}} + \kappa_{\chi} \right) - \frac{1}{2} \kappa_{\chi}  \right) \left( \frac{\mu_{\chi}}{m_P}  \right)^4 \right] \nonumber \\
	&+& 2 \, x^2 \mu_{\chi}^2 \sigma_{\chi}^2 M^2 \left( 1 + \left( 2 - \kappa_{\chi} \right) \left( \frac{\mu_{\chi}}{m_P}  \right)^2 \right)+ a \mu^2 M m_{3/2} x+ b \mu_{\chi}^2 M m_{3/2} \sigma_\chi x \nonumber \\
	&+& 2M^2 M_{S}^2 x^2+\mu_\chi^2 M_{\chi}^2 ,
\end{eqnarray}
where $\gamma _{S}=1-\frac{7\kappa _{S}}{2}+2\kappa _{S}^{2}-3\kappa _{SS}$. Here we have retained terms up to $\mathcal{O}\; (\left(\vert S \vert / m_P \right)^4, \left(\vert \chi \vert / m_P \right)^4)$ from SUGRA corrections. We turn now to the numerical analysis for non-minimal case and compute the various observables  related to inflation.

\subsection{Large $r$ solutions and Split scale SUSY}

\begin{figure}[t]
	\centering \includegraphics[width=8cm]{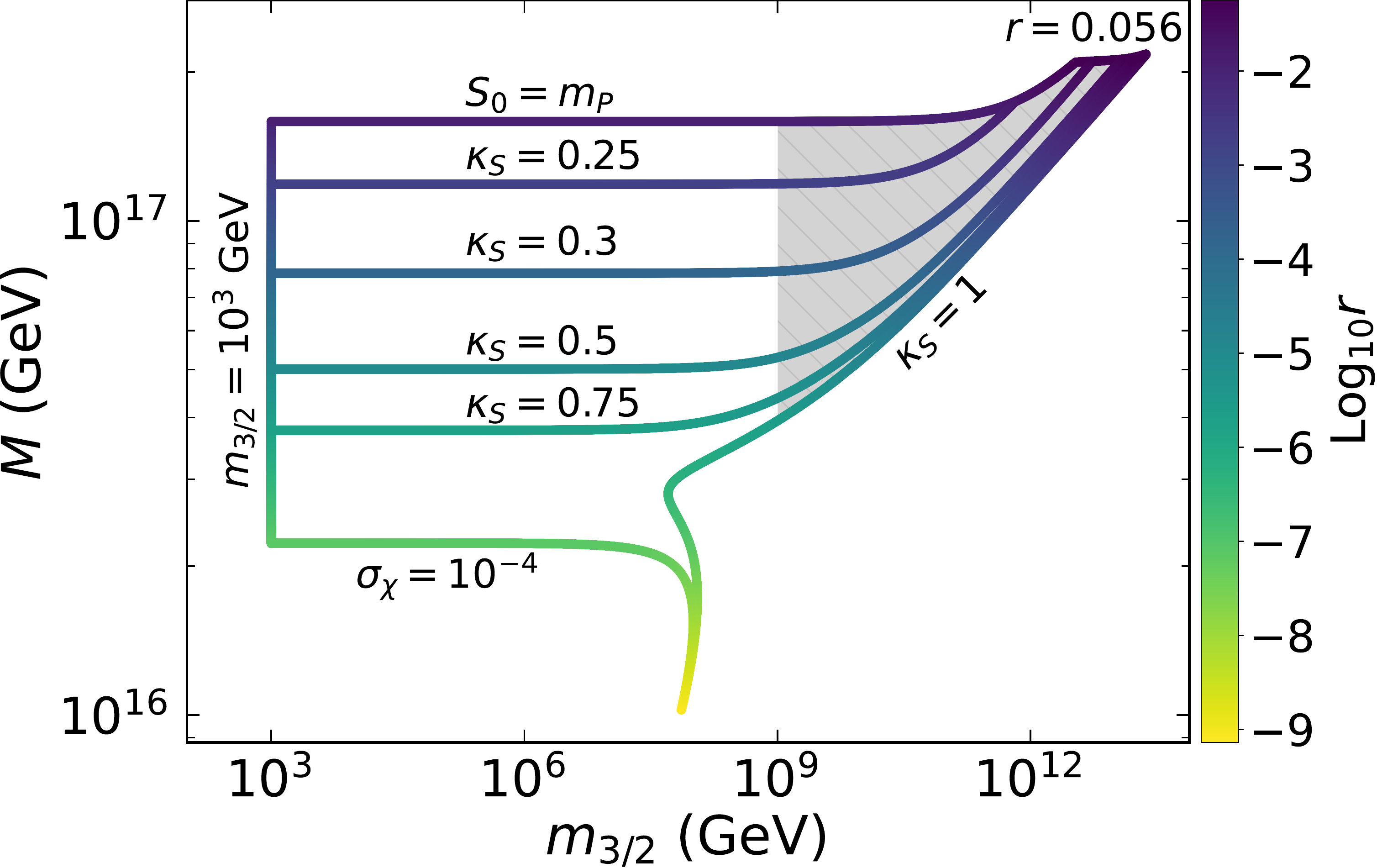}
	\centering \includegraphics[width=8cm]{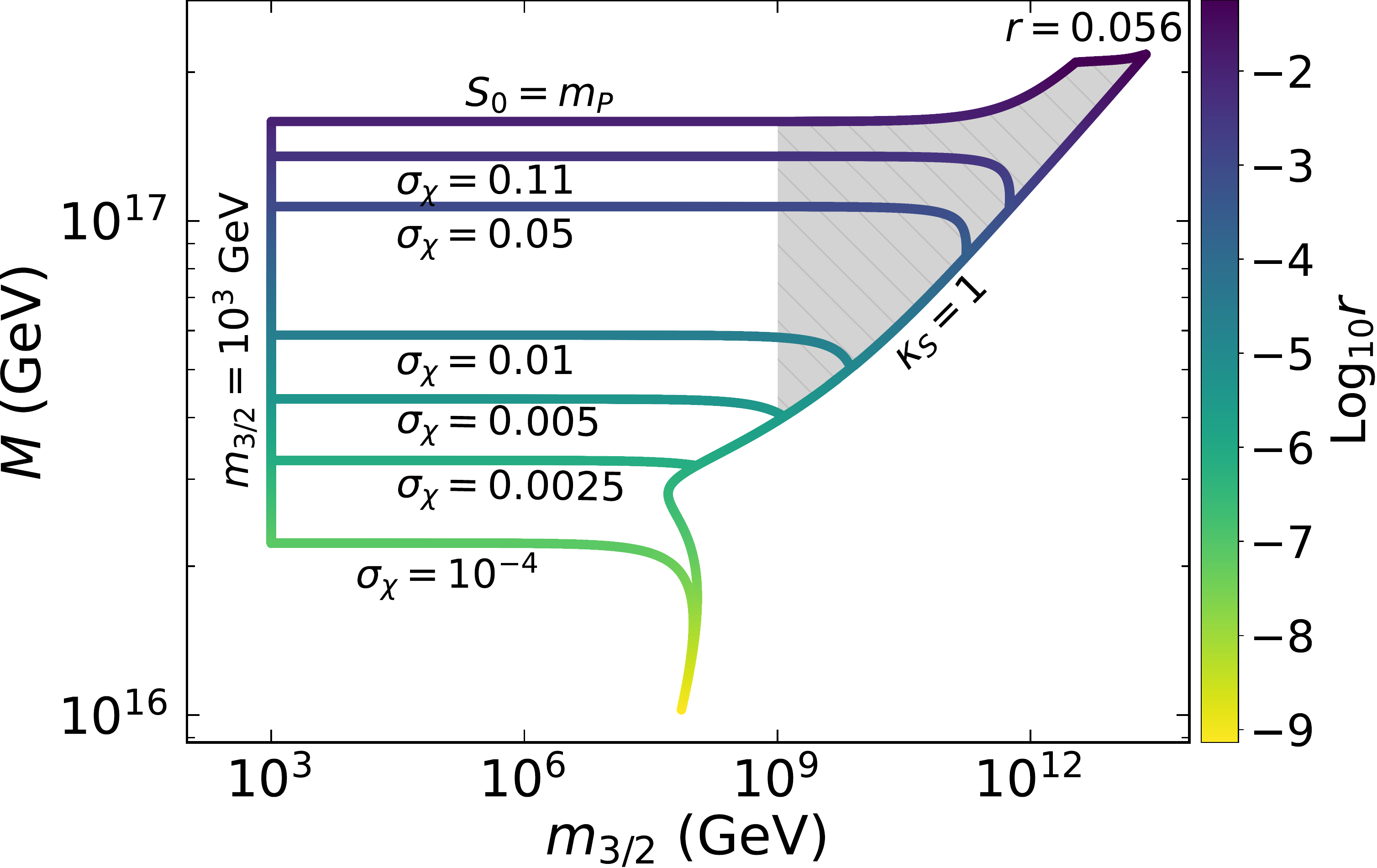}
	\caption{Variation of $\kappa_{S}$ (left panel) and $\sigma_{\chi}$ (right panel) in $m_{3/2}$-$M$ plane. The boundary curves represent $S_0 \sim m_P$, $m_{3/2} \simeq 1$ TeV, $r \simeq 0.056$, $\kappa_{S} \simeq 1$ and $\sigma_{\chi} \simeq 10^{-4}$ constraints. The color bar on the right displays the range of tensor to scalar ratio. The gray shaded region satisfies the experimental bounds on $d = 5$ proton lifetime.}
	\label{larger}
\end{figure}

\begin{figure}[t]
	\centering \includegraphics[width=8cm]{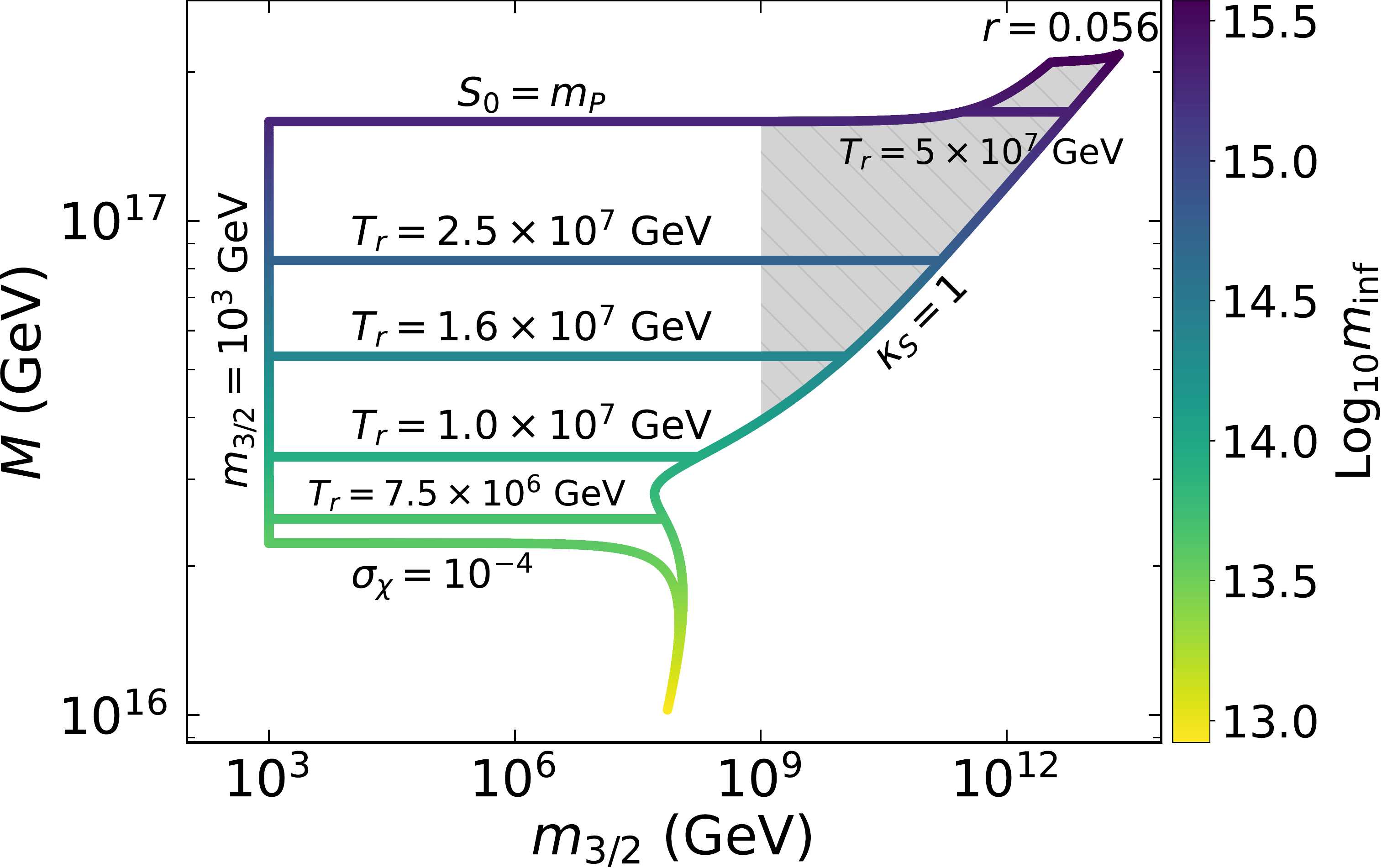}
	\centering \includegraphics[width=8cm]{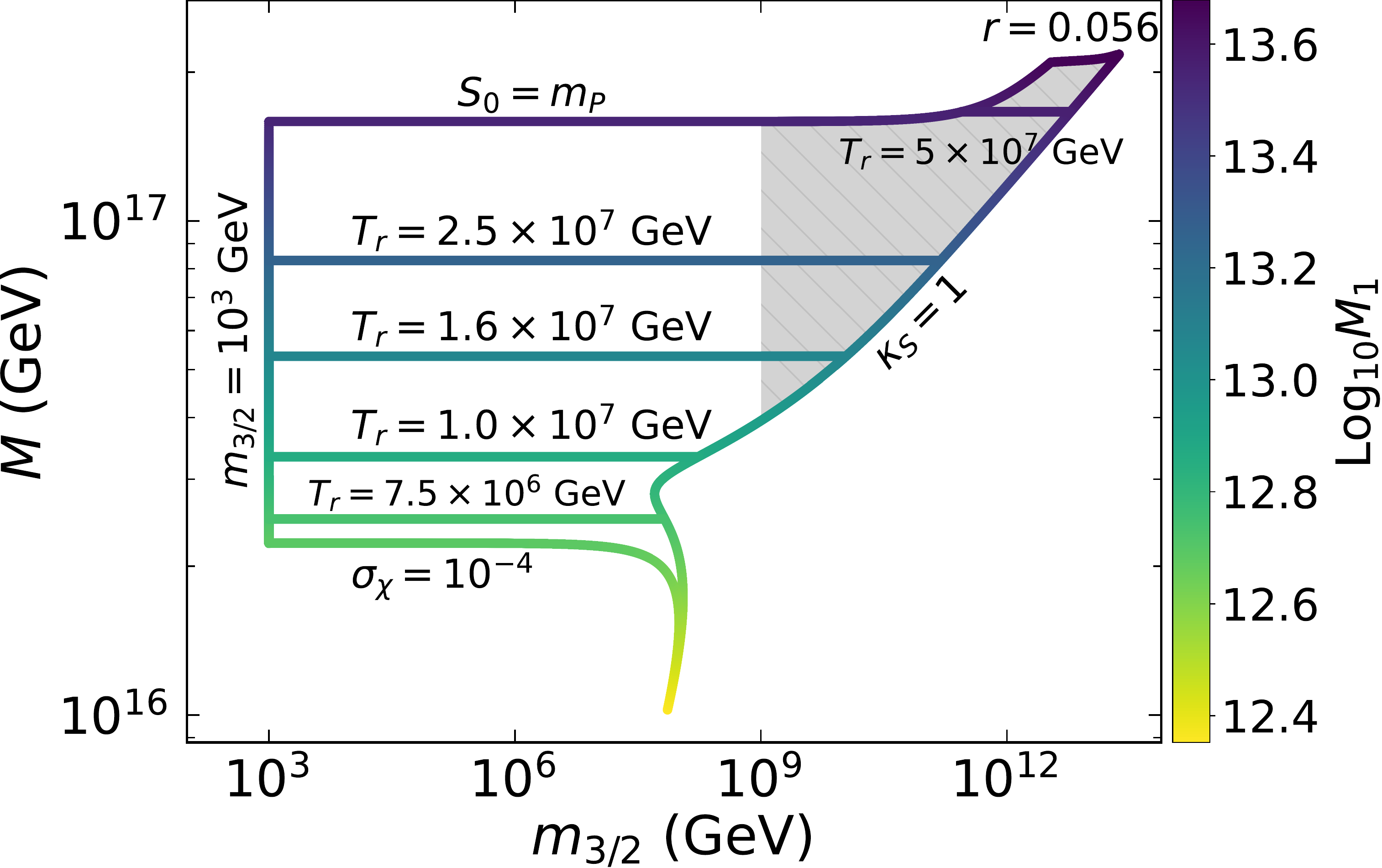}
	\caption{Variation of reheat temperature $T_r$ in $m_{3/2}$-$M$ plane. The boundary curves represent $S_0 \sim m_P$, $m_{3/2} \simeq 1$ TeV, $r \simeq 0.056$, $\kappa_{S} \simeq 1$ and $\sigma_{\chi} \simeq 10^{-4}$ constraints. The color bar displays the range of inflaton mass (left panel) and right-handed neutrino mass (right panel). The gray shaded region satisfies the experimental bounds on $d = 5$ proton lifetime.}
	\label{Trnon}
\end{figure}

The results of our numerical calculations with a non-minimal K\"ahler potential are presented in Figs. \ref{larger} and \ref{Trnon}. In obtaining these results, we have used up to second order approximation on the slow-roll parameters and we have set $\mu_{\chi}=10^{14}$ GeV, $n_{s}=0.9655$ (central value) and $a = -1$. We also set all the non-minimal couplings equal $\kappa_{S}=\kappa_{SS}=\kappa_{S\chi}=\kappa_{\chi}$, for convenience. As compared to the minimal case, the non-minimal K\"ahler potential increases the parametric space and with the addition of these new parameters, we now expect to obtain $n_s$ within the latest Planck bounds with large values of tensor-to-scalar ratio $r$. The SUGRA corrections, parametrized by $\kappa_{S}$, dominate the global SUSY potential. To keep the SUGRA expansion under control we impose $S_0 \leq m_P$. We further require that $\kappa_{S} \lesssim 1$, $\sigma_{\chi} \lesssim 1$ and restrict the tensor to scalar ratio within Planck's 2-$\sigma$ bounds, $r \lesssim 0.056$ and gravitino mass $m_{3/2} \gtrsim 1$ TeV. These constraints are shown in Figs. \ref{larger} and \ref{Trnon} which make the boundary in $m_{3/2}$-$M$ plane. The lower boundary curve is drawn for $\sigma_{\chi} \simeq 10^{-4}$. Further reduction in the value of $\sigma_{\chi}$ does not effect the inflationary predictions.

The left panel of Fig. \ref{larger} shows the variation of $\kappa_S$ whilst  the right panel shows the variation of $\sigma_{\chi}$ in the $m_{3/2}$-$M$ plane. The color bar on the right displays the range of tensor to scalar ratio $r$. For $(0.0021 \lesssim \kappa_{S} \lesssim 1)$, $(10^{-4}\lesssim \sigma_{\chi} \lesssim 1)$ and $(10^3 \lesssim m_{3/2} \lesssim 2.3 \times  10^{13} )$ GeV, we obtain $ 7.3 \times 10^{-10} \lesssim r \lesssim 0.056$, $( 1 \times 10^{16} \lesssim M \lesssim 2.1 \times 10^{17})$ GeV and $(3 \times 10^6 \lesssim T_r \lesssim 6.5 \times 10^7)$ GeV.  It can be seen that  large $r$ solutions exist  for the whole range of gravitino mass  $(1 \lesssim m_{3/2} \lesssim 10^{10})$ TeV obtained in the present  model. However, to satisfy the experimental constraints on $d = 5$ proton lifetime, we require that $m_{3/2} \gtrsim 10^6$ TeV as can be seen from Fig. \ref{fig:proton_decay}. This region which is safe from $d = 5$ proton decay favors split and high scale SUSY and is shown by the gray shaded area in Fig. \ref{larger}.
Due to the complexity of the inflationary potential, we did not include the analytical expressions for the scalar spectral index $n_s$ and tensor to scalar ratio $r$. However, the explicit dependence of $r$ on $M$ can be understood from the following approximate relation obtained by using the normalization constraint on $A_{s}$,
\begin{equation}
	r \simeq \frac{1}{360 \, \pi^2 A_{s}} \left(\frac{M}{m_P}\right)^6,
\end{equation}
which shows that $r$ and $M$ are proportional to each other and large values of $r$ are obtained for large $M$. It can readily be checked that for $M \simeq 2 \times 10^{17}$ GeV the above equation gives $r \simeq 0.04$. On the other hand, $M \simeq 2 \times 10^{16}$ GeV gives $r \simeq 4 \times 10^{-8}$. These approximate values are very close to the actual values obtained in the numerical calculations. The above equation therefore gives a valid approximation of our numerical results. The value of $r$ varies between ($7.3 \times 10^{-10} - 0.056$) which is consistent with the range ($1.0 \times 10^{16} ~ \text{GeV} \lesssim M \lesssim 2.1 \times 10^{17} ~ \text{GeV}$) shown in Fig. \ref{larger}. This shows that, small values of $r$ favor $M \sim M_{\text{GUT}}$, whereas large tensor modes require $M \gtrsim M_{\text{GUT}} \sim 2 \times 10^{17}$ GeV which in the case of a string derived model, can be identified with  the string scale. The large tensor modes can be detected by Planck and future experiments.

\subsection{Low reheating temperature and the gravitino problem}
Fig. \ref{Trnon} shows the variation of reheat temperature $T_r$ in the $m_{3/2}$-$M$ plane. The color bar in the left panel displays the range of inflaton mass $(8.35 \times 10^{12} \lesssim m_{\text{inf}} \lesssim 3.75 \times 10^{15})$ GeV whereas in right panel it displays the range of right-handed neutrino mass $(2.25 \times 10^{12} \lesssim M_1 \lesssim 4.77 \times 10^{13})$ GeV. The following two conditions
\begin{equation}
	\frac{m_{\text{inf}}}{2} \geq M_1, \qquad \qquad M_1 \geq 10 \, T_r,
\end{equation}
ensure successful reheating with non-thermal leptogenesis and are readily satisfied with the parametric range obtained in the present  model.
The reheat temperature $T_{r}$ is usually constrained by the gravitino mass $m_{3/2}$ due to gravitino overproduction. For unstable gravitinos with mass $m_{3/2}\geq 10$ TeV, the reheat temperature is almost independent of the gravitino mass, whereas for stable gravitinos, $T_{r}\leq 10^{10}$ GeV. The range of gravitino mass $(10^3 \lesssim m_{3/2} \lesssim 2.3 \times  10^{13} )$ GeV and reheat temperature $(3 \times 10^6 \lesssim T_r \lesssim 6.5 \times 10^7)$ GeV obtained in the model under consideration  naturally avoids the gravitino problem.

To avoid the rapid $d = 5$ proton decay, the model favors split and high scale SUSY for which the gravitino is short-lived and the big bang nucleosynthesis (BBN) bounds on the reheating temperature are not effective, as gravitino decays before the BBN. The gravitino decays into the lightest supersymmetric particle (LSP), neutralino $\tilde{\chi}_{1}^{0}$ for which the neutralino abundance is given by \cite{Ahmed:2021dvo}

\begin{equation}\label{eqa}
	\Omega_{\tilde{\chi}_{1}^{0}}h^{2}\simeq 2.8\times10^{11}\times Y_{3/2}\left(\frac{m_{\tilde{\chi}_{1}^{0}}}{\text{TeV}}\right),
\end{equation}
where $Y_{3/2}$ is the gravitino yield and is defined as,
\begin{equation}\label{eqb}
	Y_{3/2}\simeq2.3\times10^{-12}\left(\frac{T_{r}}{10^{10} ~ \text{GeV}}\right).
\end{equation}
Since the LSP neutralino density produced by gravitino decay should not exceed the observed dark matter (DM) relic density, choosing the upper bound of relic abundance $\Omega_{\tilde{\chi}_{1}^{0}}h^{2}=0.126$ and using equations \eqref{eqa} and \eqref{eqb}, we find a relation between the reheating temperature $T_{r}$ and neutralino mass $m_{\tilde{\chi}_{1}^{0}}$
\begin{eqnarray}\label{eqc}
	m_{\tilde{\chi}_{1}^{0}}\simeq19.6\left(\frac{10^{11} ~ \text{GeV}}{T_{r}}\right)~.
\end{eqnarray}
For gravity mediated SUSY breaking, the neutralino mass $ m_{\tilde{\chi}_{1}^{0}}\geq18$ GeV \cite{Hooper:2002nq}, which is easily satisfied for the range of reheat temperature obtained in this model.

It is worth comparing the obtained results with $SU(5)$ smooth hybrid inflation model \cite{uzubair:2015}. Note that in the limit ($\sigma_{\chi} \rightarrow 0$, $\mu_{\chi} \rightarrow 0$), the above model reduces to the smooth $SU(5)$ hybrid inflation model \cite{uzubair:2015}. With the minimal K\"ahler potential, the smooth $SU(5)$ model requires trans-Planckian field values to obtain $n_s$ and $r$ within Planck's data bounds while, in the above model, both $n_s$ and $r$ are easily obtained within Planck's data bounds with sub-Planckian field values. With a non-minimal K\"ahler potential, however, both models predict a red tilted scalar spectral index ($n_s < 1$) consistent with the Planck's latest bounds and large values of tensor-to-scalar ratio $r \lesssim 0.01$. Furthermore, in the above model of smooth $SU(5) \times U(1)_{\chi}$ hybrid inflation, the rapid $d = 5$ proton decay is avoided with split and high scale SUSY, as well as, the non-thermal leptogenesis yields a low reheat temperature $T_r \simeq 10^6$ GeV that avoids the gravitino problem for whole range of gravitino mass $m_{3/2}$.

\section{\large{\bf Summary}}\label{sum}
To summarize, we have realized smooth hybrid inflation in the supersymmetric $SU(5) \times U(1)_{\chi}$ model. The breaking of $SU(5)$ gauge symmetry during inflation dilutes the monopole density keeping it beyond the observable limit. The breaking of $U(1)_{\chi}$ symmetry leaves behind a discrete $Z_2$ symmetry, which serves as the MSSM matter parity, realizing the possibility of lightest supersymmetric particle (LSP) as a cold dark matter candidate. With $U(1)_{\chi}$ symmetry also broken during inflation, the cosmic strings are inflated away as well. We show that with a minimal K\"ahler potential, a red tilted ($n_s < 1$) scalar spectral index consistent with Planck's latest bounds requires soft SUSY breaking terms. The gravitino mass is obtained in the PeV range but is not sufficiently high to avoid the rapid $d = 5$ proton decay mediated by color-triplet Higginos. Moreover, we obtain a low reheat temperature $T_r \simeq 10^6$ GeV, although the tensor to scalar ratio remains extremely small. By employing a non-minimal K\"ahler potential, successful inflation is realized and the rapid $d = 5$ proton decay is avoided with split- and high scale SUSY. Furthermore, large tensor modes $r \lesssim 0.056$ are obtained with the non-minimal couplings $10^{-2} \lesssim \kappa_{S} \lesssim 1$ and scalar spectral index $n_s$ fixed at central value (0.9655) of Planck data bounds. The non-thermal leptogenesis yields a low reheat temperature $(3 \times 10^6 \lesssim T_r \lesssim 6.5 \times 10^7)$ GeV that avoids the gravitino problem for all range of gravitino mass $m_{3/2}$.
\vspace*{7cm}

\noindent 
{\bf Acknowledgement} 
{\it ``The  research work of the author GKL was supported by the Hellenic Foundation for
	Research and Innovation (H.F.R.I.) under the ``First Call for
	H.F.R.I. Research Projects to support Faculty members and
	Researchers and the procurement of high-cost research equipment
	grant'' (Project Number: 2251)''}.

\newpage 



\end{document}